%% file: BGW_mirage_LHC.tex
\documentclass[aps,onecolumn,nofootinbib,preprintnumbers]{revtex4}
\usepackage{amsmath,amssymb,epsfig}
\usepackage{times}
\usepackage{latexsym}
\usepackage{multirow}
\usepackage{url}
\usepackage{color}

\usepackage{graphicx}% Include figure files
\usepackage{epsfig}% Include figure files
\usepackage{amssymb}
\usepackage{color}
\input{basic.tex}

\newcommand{\comment}[1]{}

\begin{document}

\thispagestyle{empty}

\title{Mirage Models Confront the LHC:\\ I. K\"ahler-Stabilized Heterotic String Theory}
\author{Bryan L. Kaufman and Brent D. Nelson}
\affiliation{Department of Physics, Northeastern University, Boston, MA 02115, USA}
\author{Mary K. Gaillard}
\affiliation{Department of Physics and Theoretical Physics Group, Lawrence Berkeley Laboratory,
University of California, Berkeley, California 94720}
%\today

\preprint[{UCB-PTH-13/02}

\begin{abstract}
We begin the study of a class of string-motivated effective supergravity theories in light of current data from the CERN Large Hadron Collider (LHC). The case of heterotic string theory, in which the dilaton is stabilized via non-perturbative corrections to the K\"ahler metric, will be considered first. This model is highly constrained and therefore predictive. We find that much of the reasonable parameter space afforded to the model -- representing the strong dynamics of a presumed gaugino condensation in the hidden sector -- is now observationally disfavored by the LHC results. Most of the theoretically-motivated parameter space that remains can be probed with data that has already been collected, and most of the remainder will be definitively explored within the first year of operation at $\sqrt{s} = 13\,{\rm TeV}$. Expected signatures for a number of benchmark points are discussed. We find that the surviving space of the model makes a precise prediction as to the relation of many superpartner masses, as well as the manner in which the correct dark matter relic density is obtained. Implications for current and future dark matter search experiments are discussed.

\end{abstract}

\maketitle

\renewcommand{\thepage}{\arabic{page}}
\setcounter{page}{1}
\def\thefootnote{\arabic{footnote}}
\setcounter{footnote}{0}

%%%%%%%%%%%%%%%%%%%%%%%%%%%%%%%%%%%%%%%%%%%%%%%%%%%%%%%%%%%%%%%%%%%%%%
\section{Introduction}

The Large Hardron Collider (LHC) has now ushered in the long-awaited data era for physics beyond the Standard Model. The recent discovery of a resonance consistent with the Higgs boson of the Standard Model~\cite{:2012gu,:2012gk}, and the observation of the rare decay $B_s \to \mu^+ \mu^-$~\cite{:2012ct}, has triumphantly affirmed the Standard Model while providing crucial clues as to what may, and may not, be the next great theory of nature. Supersymmetry has long held pride of place among such postulated theories, and while the recent results are generally in conflict with very low superpartner masses, the notion that supersymmetry is relevant for physics at the electroweak scale is still very much a compelling paradigm. 

As most sensible theories of low-energy physics arising from string compactification contain supersymmetry in four dimensions, the on-going LHC data continues to raise the stakes for the notion that a meaningful `string phenomenology' exists. It is entirely appropriate for the string theory community to pause and reflect on what, if anything, the LHC data is saying about the construction of low-energy particle physics from string compactification. In this regard, `top-down' models are a good place to begin asking questions about how LHC data is shaping the theoretical consensus on how supersymmetry is broken and then transmitted to the fields of the observable sector. Such questions require a model context to be sensible scientific questions.

String-theoretic models must have stabilized moduli to make {\em bona fide} statements about supersymmetry breaking, and hence the superpartner spectrum. One is thus restricted to only a few well-studied examples in the string phenomenology literature. Here we will focus on the class of models which give rise to the so-called `mirage pattern' of gaugino masses~\cite{Choi:2007ka}, in which the ratios of the soft supersymmetry-breaking gaugino masses at the electroweak scale are governed by the approximate formula
\begin{equation} M_1\,:\,M_2\,:\,M_3 \, = \, (1.0+0.66\alpha)\, : \,
(1.93+0.19\alpha) \, : \, (5.87-1.76\alpha)\, .
\label{mirage_ratios} 
\end{equation}
The parameter $\alpha$ is determined by the model-dependent dynamics which stabilize the relevant moduli in the theory, and it is indirectly measurable with sufficient LHC data through its influence on various collider observables~\cite{Altunkaynak:2009tg}. For example, when $\alpha \simeq 2$ gaugino soft masses are nearly equal at the electroweak scale. Such a compressed gaugino mass spectrum may allow a gluino to exist in the LHC~data at a mass significantly smaller than the nominal limit coming from direct searches~\cite{Baer:2007uz,Dreiner:2012gx}. For most semi-realistic string models -- and certainly for the model considered in this paper -- such values are unlikely to arise, yet substantial departures from the ratios predicted by unified theories such as minimal supergravity (mSUGRA)~\cite{Chamseddine:1982jx,Nath:1983aw,Hall:1983iz} are nevertheless likely. We will refer to any string-motivated model of supersymmetry breaking whose gaugino sector follows the pattern in~(\ref{mirage_ratios}) as a `mirage model', though this term will include theories with widely differing patterns of soft supersymmetry-breaking scalar masses.

There are many reasons to begin a study of string phenomenology in the LHC~data era with such models. As mentioned above, the mirage pattern in the gaugino sector is known to allow for lighter gluino masses relative to squark masses than in more constrained models, given the current direct search strategies at the LHC. More importantly for our purposes, the mirage pattern in the gaugino sector is a surprisingly robust prediction of a diverse array of string compactifications. The reason for this ubiquity can be identified in the rather simple properties all such mirage models share: (1) the modulus which determines the gauge couplings in the observable sector is the last modulus to be stabilized, (2) this stabilization involves gaugino condensation in a hidden sector, and (3) the non-canonical kinetic term for this modulus is effectively altered from its tree-level form via some sort of non-perturbative effect, which is tuned to result in vanishing vacuum energy in the eventual minimum. It is the last property that contains the model-dependence and which distinguishes one manifestation of the mirage pattern from another.

Finally, mirage models represent some of the most thoroughly studied effective supergravity theories within the context of string phenomenology. The models which form the focus of this paper -- K\"ahler-stabilized heterotic string compactifications -- were the first such models to be constructed and remain among the most predictive due to their relatively small number of free parameters~\cite{Gaillard:2007jr}. The three conditions identified above, in the context of heterotic strings, are precisely the ingredients of what Casas referred to as the `generalized dilaton domination' scenario~\cite{Casas:1996zi}. The model, first constructed by Bin\'etruy, Gaillard and Wu (henceforth 'BGW')~\cite{Binetruy:1996xja,Binetruy:1996nx}, can therefore be thought of as a concrete top-down manifestation of a bottom-up phenomenological paradigm. We will support these observations in the following section, in which the broad theoretical outlines of the BGW~scneario are laid out. In Section~\ref{scan} we put the BGW~model to the test in confronting data, especially the impact on the parameter space of the theory from the observation of a Standard Model-like Higgs boson with a mass near 126~GeV, and the constraint from the WMAP~satellite on the relic density of stable neutralinos.
In Section~\ref{LHC} we consider the limits from direct searches for superpartners at the~LHC on that part of the theory space which still remains. Gluinos with masses below 800-900~GeV are excluded in the BGW~model, primarily from searches for same-sign dileptons with accompanying b-tagged jets.
In general, we find that the model is highly constrained, with the remaining viable parameter space predicting an ensemble of electroweak charginos and neutralinos with high Higgsino content in the mass range of 300-700~GeV. In Section~\ref{futureLHC} we comment on the prospects for discovery of superpartners at the LHC at post-shutdown center-of-mass energies of $\sqrt{s} = 13\,{\rm TeV}$. We also comment on the likelihood of measuring the presence of neutralino dark matter in current and future dark matter detection experiments. 

%%%%%%%%%%%%%%%%%%%%%%%%%%%%%%%%%%%%%%%%%%%%%%%%%%%%%%%%%%%%%%%%%%%%%%
%\section{Mirage Models in General}
%\input{input_mirage.tex}

%%%%%%%%%%%%%%%%%%%%%%%%%%%%%%%%%%%%%%%%%%%%%%%%%%%%%%%%%%%%%%%%%%%%%%
\section{The BGW Model}
\label{bgw}

The BGW model is motivated from orbifold compactifications of heterotic string theory, though many of its features would be equally applicable to smooth Calabi-Yau compactifications. The moduli sector of the theory will here be confined to three K\"ahler moduli and a single dilaton field. The low-energy supergravity theory will be assumed to respect an $SL(2,\mathbb{Z})$ symmetry operating upon the K\"ahler moduli. We will refer to this symmetry as simply `modular invariance' when describing the effective Lagrangian. Such modular transformations are classical symmetries of the supergravity theory, but are anomalous at the quantum level. The invariance is restored by a string-derived Green-Schwarz counterterm~\cite{Green:1984sg,Derendinger:1991hq,LopesCardoso:1991zt,LopesCardoso:1992yd} and possible stringy threshold corrections to gauge couplings~\cite{Kaplunovsky:1995jw}.

Moduli stabilization is to be achieved in the BGW model via confinement of hidden sector gauge groups. More than one such group is accommodated in the model, as are possible matter condensates for fields charged under the confining gauge groups. K\"ahler moduli will be stabilized at self-dual points, while the dilaton is minimized in such a way as to achieve a weak-coupling solution to the scalar potential and vanishing vacuum energy simultaneously. This is accomplished by introducing non-perturbative corrections to the dilaton action arising from string instanton effects. These corrections are treated in a phenomenological manner, and result in a modification of the K\"ahler metric for the dilaton in the component Lagrangian. As stated above, this can be thought of as an implementation of the generalized dilaton domination scenario of Casas, which will necessarily give rise to the mirage pattern of gaugino masses.

\subsection{Walk-Through of the Model}

The goal of this section is to lay out the features of the BGW model and specify notation so as to identify the free parameters of the theory. Additional background and detail are provided in the references given throughout this section. We note that we will display terms in the effective superspace Lagrangian using the formalism of K\"ahler U(1) superspace~\cite{Binetruy:2000zx}. Most intuition as to the resulting component Lagrangian from the formalism of Wess and Bagger~\cite{WB} continues to hold; a useful primer into the differences can be found in the brief appendix to reference~\cite{Gaillard:2007jr}.

The choice of K\"ahler U(1) superspace is motivated in large part because it naturally incorporates the string theory dilaton into a real, linear multiplet~\cite{Binetruy:1991sz}, in which form the implementation of non-perturbative effects arising from string theory is made far more transparent. Since these effects form the crux of the BGW stabilization mechanism, the linear multiplet formulation is highly preferable to the familiar chiral formulation. At the leading order the two formalisms are related by the simple superfield identification
\begin{equation} L = \frac{1}{S+\oline{S}} 
\label{treeLSdual}
\end{equation}
though this identification fails to be satisfactory at higher loop level.
The lowest component $\ell$ of the superfield $L$ is then
the dilaton and the vacuum expectation value (VEV) of the lowest component $\ell =
L|_{\theta=\bar{\theta}=0}$ determines the (universal) gauge coupling constant
$g_{\STR}$ at the scale $\Lambda_{\STR}$ via the relation
\begin{equation} \frac{g_{\STR}^2}{2} = \lang \ell \rang \; . 
\label{gbare2}
\end{equation}
This quantity represents the string loop expansion parameter.
Therefore string theory information from higher loops is more
naturally encoded in terms of this set of component fields. 

Stabilization will occur through non-perturbative corrections to the dilaton action motivated by string instanton calculations. Such corrections can be represented as a modification to the K\"ahler potential $k(L)$ for the dilaton, though in the K\"ahler U(1) superspace formulation it is the modifications to the action, via the kinetic superspace Lagrangian, that should be taken as the fundamental quantity. We can therefore parameterize the corrections by two (related) functions of the linear multiplet
\begin{equation}
\Lag_{\rm KE} = \superint\,E \[-2 + f(L)\], \quad k(L) = \ln\,L +
g(L)\, ,
\label{lke}
\end{equation}
where $f(L)$ and $g(L)$ are related by the differential equation
\begin{equation}
L\frac{\diff g(L)}{\diff L}\,=\, -L\frac{\diff f(L)}{\diff
L}\,+\,f(L) \, .
\label{lcond} 
\end{equation}
Note that the tree level dilaton K\"ahler potential is the straightforward analog to that of the chiral formulation $K=-\ln(S+\oline{S})$. Including the three K\"ahler moduli $T^I$ we have the complete K\"ahler potential for the geometrical moduli
\begin{equation} K = \ln(L) + g(L) - \sum_I\ln\(T^I + \oline{T}^I\) \, .
\label{Kmoduli}
\end{equation}
As we wish to obtain the classical limit at weak
coupling, we impose a further boundary condition at
vanishing coupling that $g(L=0)\,=\,0$ and $f(L=0)\,=\,0$. In the presence of these nonperturbative effects the relationship
between the dilaton and the effective field theory gauge coupling is
modified from the relation in~(\ref{gbare2}) to
\begin{equation} \frac{g_{\STR}^2}{2}=\lang \frac{\ell}{1+f\(\ell\)} \rang \, .
\label{gnew} 
\end{equation}

The corrections to the dilaton action will take the form of a sum over instanton corrections, with undetermined coefficients $A_n$~\cite{Shenker:1990uf,Polchinski:1994fq,Silverstein:1996xp,Antoniadis:1997zt}
\begin{equation}
f(L) = \sum_{n} A_n (\sqrt{L})^{-n} e^{-B/\sqrt{L}}\; .
\label{nonpertsum}
\end{equation}
For simplicity we have taken a single overall normalization for the argument of the exponent, but this can easily be relaxed. It is an important feature of~(\ref{nonpertsum}) that these string instanton effects scale like $e^{-1/g}$ (when we use $\ell \sim g^2$) and are thus stronger than analogous nonperturbative effects in field theory which have the form $e^{-1/g^2}$.They can therefore be of significance even in cases where the effective four-dimensional gauge coupling at the string scale is weak~\cite{Banks:1994sg}.

Subsequent phenomenology is largely insensitive to the precise values of the coefficients in the series~(\ref{nonpertsum}). We will therefore allow ourselves the freedom to assume that the coefficients can be tuned so as to achieve the three necessary conditions of the minimum: (1) that a minimum at finite $\lang \ell \rang$ is achieved, (2) that the dilaton potential vanishes in the vacuum at that value, and (3) that the string coupling given by~(\ref{gnew}) satisfies $g_{\STR}^2 = 1/2$ at the minimum.
In practice, it is sufficient to truncate this sum to only the first two terms
\begin{equation} f(L) = \(A_0 + \frac{A_1}{\sqrt{L}}\)e^{-B/\sqrt{L}} \, ,
\label{ourf} 
\end{equation}
which allows a simultaneous solution to be obtained for $\lang L \rang = \ell$ and two out of the three constants $A_0$, $A_1$ and $B$, with the third being fixed beforehand.

The other ingredient for our dilaton stabilization mechanism is provided by non-perturbative field theoretic contributions arising from gaugino condensation in the hidden sector of the theory. We will allow for a general hidden sector group
\begin{equation}
\mathcal{G}_{\rm hidden} = \prod_{a=1}^n\mathcal{G}_a\otimes U(1)^m
\, .
\label{Ghidden}
\end{equation}
where the label $a$ counts non-Abelian factors in the hidden sector. Gaugino condensates are represented in superspace by a composite field operator $U_a$~\cite{Veneziano:1982ah}, which denotes a $\mathcal{G}_a$-charged gauge condensate chiral superfield
\begin{equation} U_a\simeq {\Wc_a\Wa^a} \, ,
\label{Ucondef} 
\end{equation}
where the lowest component of $u_a = U_a\lowest$ involves the gaugino bilinear $\lambda_a \lambda_a$. We also wish to add the possibility of
matter charged under the condensing group. From
our experience in QCD we generally expect states charged under the
strong group to experience confinement and form composites. We will
represent these by the composite field operators~\cite{Taylor:1982bp,Lust:1990zi}
\begin{equation} \Pi_a^\alpha \simeq
\prod_{i} \(\Phi_i^{(a)}\)^{n_i^{\alpha,(a)}},\label{Picondef}
\end{equation}
where the product involves only those fields $\Phi_i^{(a)}$ charged
under the confined group $\mathcal{G}_a$. In~(\ref{Picondef}) the
label $\alpha$ is a species index for the matter condensates, each
of which may consist of different component fields labeled by the
integers $n_i^{\alpha,(a)}$. Note that the canonical mass dimension
of this operator $\Pi_a^\alpha$ is given by
\begin{equation}
{\rm dim}\(\Pi_a^\alpha\)\equiv d_a^{\alpha} = \sum_{i}
n_i^{\alpha,(a)} \label{dimpi}
\end{equation}

In K\"ahler U(1) superspace the effective Lagrangian describing these condensates takes the Veneziano-Yankielowicz-Taylor form~\cite{Taylor:1985fz,Binetruy:1989dk}
\begin{equation} \Lag_{\rm VYT} = \frac{1}{8} \superint\,\frac{E}{R}\sum_a
U_a\[b'_a\ln(e^{-K/2}U_a) + \sum_\alpha b^\alpha_a\ln\Pi_a^\alpha\]
+ {\rm h.c.}\; , \label{VYT} \end{equation}
where there are now two separate coefficients $b'_a$ and $b_a^{\alpha}$ which must be determined for each condensing group $\mathcal{G}_a$. These are obtained by matching the anomalies of the effective theory to those of the underlying theory. The Lagrangian~(\ref{VYT}) has the correct anomaly structure under K\"ahler $U(1)$, R-symmetry, conformal transformations, and modular (T-duality) transformations, provided the conditions
\begin{equation} b'_a = \frac{1}{8\pi^2}\(C_a - \sum_iC_a^i\) ,\;\;\;\; \qquad
b^\alpha_a = \sum_{i\in\alpha}\frac{C^i_a}{4\pi^2 d_a^\alpha}\; ,
\label{cond} \end{equation}
are satisfied~\cite{Gaillard:1992bt}. In~(\ref{cond}) $C_a$ and $C_a^i$ are the quadratic Casimir operators for the gauge group $\mathcal{G}_a$ in the adjoint representation
and in the representation of the matter fields $Z^i$ charged under
that group, respectively. 
Note the important property that when $d_a^{\alpha}=3$ for all
$\Pi$'s charged with respect to the condensing group $\mathcal{G}_a$,
we have the identity
\begin{equation} b'_a + \sum_{\alpha} \baal = b_a = \frac{1}{8\pi^2} \( C_a - \frac{1}{3} \sum_i C_a^i \)\, , \label{ba}
\end{equation}
with $b_a$ being the beta-function coefficient associated
with the coupling for the group $\mathcal{G}_a$. We will make the assumption that $d_a^{\alpha}=3$ in what follows below. The conventions in~(\ref{ba}) imply that a
group $\mathcal{G}_a$ with $b_a > 0$ will flow to strong coupling in
the infrared.\footnote{To recover the `standard'
conventions of (for example) Martin and Vaughn~\cite{Martin:1993zk}
one must take $b_a \to -(2/3) b_a|_{\rm MV}$.}

The composite chiral superfields $\Pi^\alpha_a$ are
invariant under the nonanomalous symmetries, and may be used to
construct an in\-var\-i\-ant
su\-per\-po\-ten\-tial~\cite{Taylor:1982bp,Gaillard:2003gt}
\begin{equation}
{\Lag}_{\rm pot}=\frac{1}{2}{\superint}\frac{E}{R}{e^{K/2}}W\(
\Pi^{\alpha}, \, T^{I} \) + \hc\;. 
\label{Lagpot}
\end{equation}
We will adopt the simplifying assumption~\cite{Binetruy:1996nx} that for fixed $\alpha$, $\baal\neq 0$ for only one value of $a$. In other words, we assume that each matter condensate is made up fields charged under only one of the confining groups. This is not a necessary requirement, but it will make the phenomenological analysis of the model much easier to perform. We next assume that there are no unconfined matter fields charged under the confined hidden sector gauge groups, and ignore possible dimension-two matter condensates involving vector-like pairs of matter fields. This allows a simple factorization of the superpotential of the form
\begin{equation}
W\( \Pi^{\alpha}, \, T^{I} \) = {\sum_{\alpha}}{W_{\alpha}}\(T\)
{\Pi}^{\alpha} ,
\end{equation}
where the functions $W_{\alpha}$ are given by
\begin{equation}
W_{\alpha}\(T\)=c_{\alpha}\prod_I \[\eta \(T^{I}\)\]^{2\(
q^{\alpha}_{I} -1\)}. 
\label{Walpha}
\end{equation}
Here $q^{\alpha}_{I}=\sum_{i} n_i^{\alpha}q_{i}^{I}$ is the
effective modular weight for the matter condensate. It will not be important for the phenomenology which follows. The Yukawa coefficients $c_{\alpha}$ are {\it a priori} unknown variables, and the function $\eta(T^I)$ is the classical Dedekind eta-function, which endows the superpotential with the proper transformation property under $SL(2,\mathbb{Z})$ modular transformations. 

Such transformations are classical symmetries of the supergravity theory, but are anomalous at the loop level. This anomaly is canceled by a combination of the Green-Schwarz (GS) mechanism and threshold corrections to the gauge kinetic functions. The GS mechanism is represented by a superspace interaction term between the dilaton and the K\"ahler moduli given by
\begin{equation}
{\Lag}_{\rm GS}=-b_{\GS} {\superint}\, E\, L\, {\sum_I} \ln\(T^I +
\oline{T}^I\) 
\label{LGS}
\end{equation}
where the coefficient $b_{\GS}$ is normalized as $b_{\GS} = C_{\GS}/8\pi^2$. The Green-Schwarz coefficient $C_{\GS}$ is a calculable parameter of the compactification which, for example, will satisfy $C_{\GS} \leq 30$ for heterotic string theory compactified on orbifolds~\cite{Giedt:2001zw}.  Threshold corrections are represented by an additional term in the effective Lagrangian given by
\begin{equation}
\Lag_{\rm th} =
-\sum_{I}\frac{1}{64\pi^2}\superint\,\frac{E}{R}\[\sum_a b_{a}^I U_{a}\] \ln\eta^2(T^I)  + {\rm h.c.}
\label{Lthresh} \end{equation}
The coefficients $b_a^I$ are themselves completely determined by the value of $b_{\rm GS}$ in~(\ref{LGS}) and by the charges and modular weights of the confined fields; they do not appear explicitly in the expression for the potential given below.
Modular invariance will yield a solution in which all three K\"ahler moduli are stabilized at one of the self dual points $\langle t^I \rangle = 1$ or  $\langle t^I \rangle = e^{i\pi/2}$, where the associated auxiliary fields $F_{T^I}$ vanish in the vacuum. Thus, the BGW model is an example of the generalized dilaton domination scenario of Casas.

\subsection{Moduli Stabilization and Supersymmetry Breaking}
\label{modstab}

The dynamical degrees of freedom associated with the composite fields~(\ref{Ucondef}) and~(\ref{Picondef}) acquire masses larger than the condensation scale $\Lambda_a$, and may be integrated out~\cite{Wu:1996en}. This results in an effective theory constructed as described in~(\ref{VYT}) with the composite fields taken to be nonpropagating. We can therefore find an expression for the gaugino condensates by solving the equations of motion for the auxiliary fields $F_{U_a} +\oline{F}_{\oline{U}_a}$~\cite{Binetruy:1996nx}
\begin{equation}
u_a^2 = e^{-2{\frac{b'_a}{b_a}}}e^{K}
e^{-\frac{\(1+f\)}{{b_a}\ell}}e^{-\frac{b_{\GS}}{b_a}
  {\sum_I}\ln\(t^I +
\oline{t}^I\)}{\prod_I}
  \left|{\eta}\(t^{I}\)\right|^{\frac{4\(b_{\GS}-b_{a}\)}{b_a}}
{\prod_{\alpha}}\left|\baal/4c_{\alpha}\right|^{-2
{\frac{b_{a}^{\alpha}}{b_a}}} \, . 
\label{uasq}
\end{equation}
The explicit expression for the condensate value is sufficiently complicated to warrant some commentary before proceeding. First we note that the expression is dimensionless by construction -- physical masses and scales will be shown explicitly below with the Planck mass restored. The third exponential term is the familiar one from field-theoretic arguments
\begin{equation} \lang \lambda_a \lambda_a \rang \sim M^3_{\PL}
e^{-1/b_a g_a^2(M_{\PL})}\; . 
\label{lamlam} 
\end{equation}
The product of Dedekind functions with the fourth exponential term provides an overall transformation property for the superpotential consistent with target space modular invariance~\cite{Nilles:1990jv,Binetruy:1990ck,Font:1990nt,Nelson:2002fk}. The final product is over the matter condensates in the theory (labeled by $\alpha$), and involves the Yukawa couplings $c_{\alpha}$ of~(\ref{Walpha}). This is consistent with the exact beta-function of Shifman and Vainshtein~\cite{Shifman:1986zi}, where the final term represents the contribution from wave-function renormalization~\cite{Amati:1988ft}. 

The scalar potential for the dilaton can now be identified in terms of the gaugino condensate factors 
\begin{equation} V = \frac{1}{16\ell^2} \(1+\ell\frac{\diff
  g}{\diff \ell} \)\left|\sum_a\(1+b_a\ell\)u_a\right|^2 -\frac{3}{16}\left|\sum_a b_a u_a\right|^2 \, ,
\label{Vdil}
\end{equation}
and the vacuum expectation value of the supergravity auxiliary field determines the gravitino mass as
\begin{equation} m_{3/2} = \frac{M_{\PL}}{4} \langle | \sum_a b_a u_a | \rangle \, .
\label{mgrav}
\end{equation}
In practice, even very small differences in the value of $b_a$ between the various confining gauge groups can cause large differences in the scale of gaugino condensation, and hence very different contributions to the scale of supersymmetry breaking represented by the gravitino mass in~(\ref{mgrav})~\cite{Gaillard:1999et}. It is therefore sufficient to consider the largest value of $b_a$ among all of the confining groups, which we hereafter refer to as $b_+$. Then we have the expressions
\begin{equation} m_{3/2} = \frac{M_{\PL}}{4}b_+ u_+ \, , \qquad \Lambda_{\rm cond} = M_{\PL} \(u_+\)^{1/3}
\label{mgravplus}
\end{equation}
and the dilaton potential can be minimized with the overall scale, determined by $|u_+|^2$, factored out of the expression in~(\ref{Vdil}).

The above parameter space can be simplified greatly by assuming that
all of the matter in the hidden sector which transforms under a
given subgroup $\mathcal{G}_a$ is of the same representation, such
as the fundamental representation. This is not unreasonable given
known heterotic string constructions. In this case the sum of the
coefficients $\baal$ over the number of condensates can be replaced
by one effective variable
\begin{eqnarray}
{\sum_{\alpha}}\baal\longrightarrow\baaleff\, ;  & \qquad
\baaleff={N_c}b_{a}^{\rm rep}\, . \label{baaleff}
\end{eqnarray}
In the above equation $b_{a}^{\rm rep}$ is proportional to the
quadratic Casimir operator for the matter fields in the common
representation and the number of condensates, $N_c$, can range from
zero to a maximum value determined by the condition that the gauge
group presumed to be condensing must remain asymptotically free.
The variable $\baal$ can then be eliminated in~(\ref{uasq}) in
favor of $\baaleff$ provided the simultaneous redefinition
$c_{\alpha}\longrightarrow\(c_{\alpha}\)_{\rm eff}$ is made so as to
keep the final product in~(\ref{uasq}) invariant. Combined with the
assumption of universal representations for the matter fields, this
implies
\begin{equation}
\caleff \equiv
{N_c}\({\prod_{\alpha=1}^{N_c}}c_{\alpha}\)^\frac{1}{N_c} \, .
\label{caleff}
\end{equation}

\subsection{Soft Supersymmetry Breaking in the Observable Sector}

The massless spectrum of a linear multiplet contains no auxiliary fields. Thus, there is no analog to the highest component $F^S$ of the chiral dilaton formulation to serve as an order parameter for supersymmetry breaking. Supersymmetry breaking soft-terms for the fields of the observable sector (here assumed to be those of the MSSM) must be read directly from the component Lagrangian itself~\cite{Binetruy:1996nx,Gaillard:1999et,Gaillard:1999yb}. It is far more instructive, however, to translate these results into the familiar chiral language of, for example, Brignole et al.~\cite{Brignole:1993dj}. This was the approach taken in~\cite{Kane:2002qp}, and we here reproduce some of the discussion presented more fully there.

The basic idea is to note that the effect of the non-perturbative corrections to the dilaton action is to modify the effective Kahler metric for the chiral dilaton away from its tree level value of $\lang (K_{s\bar{s}}^{\rm
tree})^{1/2} \rang = \lang 1/(s+\bar{s}) \rang = g_{\STR}^{2}/2
\simeq 1/4$. We can parameterize this departure via the quantity
\begin{equation}
a_{\rm np} \equiv \(\frac{K_{s\bar{s}}^{\rm
tree}}{K_{s\bar{s}}^{\rm true}}\)^{1/2} \, .
\label{acond}
\end{equation}
As with the phenomenological approach of Brignole et al., we can now demand that supersymmetry breaking occur, with $\lang F^S \rang \neq 0$, while simultaneously demanding that $\lang V \rang =0$. In such circumstances, in which the dilaton is the sole source of supersymmetry breaking, a simple relation must exist between the value of $F^S$ and the gravitino mass in the vacuum
\begin{equation}
F^{S} = \sqrt{3} m_{3/2} (K_{s\bar{s}})^{-1/2} =
\sqrt{3}m_{3/2}a_{\rm np}(K_{s\bar{s}}^{\rm tree})^{-1/2} , .
\label{FS}
\end{equation}
It was the simple observation that vanishing vacuum energy would require $a_{\rm np} \neq 1$ that gave birth to the generalized dilaton domination scenario~\cite{Casas:1996zi}.

The BGW~model provides an explicit expression for $a_{\rm np}$ in terms of the non-peturbative correction in~(\ref{nonpertsum}). Relating the chiral and linear multiplet formulations, in the presence of the nonperturbative effects given in~(\ref{lcond}), yields the following expressions for the derivatives of the K\"ahler potential in the vacuum
\begin{equation}
\lang K_s \rang = -\ell \qquad \qquad \lang K^{\rm true}_{s\bar{s}}\rang =
\frac{\ell^2}{1+\ell g'(\ell)} \, . 
\label{translate}
\end{equation}
To make this relation more specific, we can choose to minimize~(\ref{Vdil}) for a single condensate $b_a \to b_+$ using the parameter set
\begin{equation}
A_0 = 8.9 \qquad A_1 = -4.5 \qquad B= 0.75  
\label{parameters}
\end{equation}
in~(\ref{ourf}). This yields a solution for which $\lang f(\ell) \rang \simeq 0$ and thus $\lang \ell \rang \approx g_{\STR}^{2}/2$ as it would be in the perturbative limit. In this case, using the fact that~(\ref{Vdil}) must vanish in the vacuum, we can write the parameter $a_{\rm np}$ explicitly as
\begin{equation}
a_{\rm np}=\sqrt{3}\frac{\frac{g_{s}^{2}}{2}b_{+}}{1+
\frac{g_{s}^{2}}{2}b_{+}} \, . 
\label{aBGW}
\end{equation}
We note that since $b_+ \sim \order (0.1)$, the quantity $a_{\rm np}$ is generally much less than unity.

Tree-level soft supersymmetry breaking terms for dimension-one quantities (gaugino masses and trilinear A-terms) arise from the dilaton via $F^S$, while those for the scalar masses get a direct contribution from the gravitino mass alone. However, from~(\ref{FS}) it is clear that in the vacuum we have
\begin{equation} \frac{F^S}{m_{3/2}} \sim a_{\rm np} \ll 1 \, , 
\end{equation}
and thus we should expect loop-corrections to be important for both the gaugino masses and the scalar triliear couplings. Full one-loop expressions for soft supersymmetry breaking in general supergravity effective theories were computed in~\cite{Gaillard:1998bf,Gaillard:1999yb,Gaillard:2000fk,Binetruy:2000md}. Here we collect only what we need for the generalized dilaton domination paradigm. 

Gaugino masses for the observable sector gauge groups ($a = SU(3),\,SU(2),\,U(1)_Y$) are given by
\begin{equation}
M_{a} = \frac{g_{a}^{2}\(\mu_R\)}{2} \left[ -3 b_{a}m_{3/2} +\( 1 -
b_{a}' K_s \) F^{S} \right] \, ,
\label{gauginomass}
\end{equation}
where $\mu_R$ is the boundary condition scale, which we take to coincide with the cutoff scale which can be taken to be the scale at which the supergravity approximation breaks down. For calculational purposes we will take the common approach of treating this scale to be the grand unification scale at which gauge couplings are approximately unified. The first term in~(\ref{gauginomass}) is the contribution from the superconformal anomaly~\cite{Gaillard:1999yb,Bagger:1999rd}. The second term is the universal contribution from the dilaton (plus the one-loop correction to that universal piece). From~(\ref{FS}) we see that this model is indeed a mirage model, with the value of the parameter $\alpha$ given, in the conventions of Choi et al.~\cite{Choi:2005uz}, by
\begin{equation} \alpha = \frac{1}{\sqrt{3}\ln\(M_{\PL}/m_{3/2}\)a_{\rm
np}}  \, . 
\label{alphaBGW}
\end{equation}
The soft supersymmetry-breaking trilinear scalar couplings and scalar mass-squareds are given by a combination of bulk supergravity contributions and superconformal anomaly contributions
\begin{eqnarray}
A_{ijk} &=& -\frac{K_s}{3}F^S
+ \frac{1}{3} \gamma_{i}m_{3/2} + {\rm cyclic}(ijk) \nonumber \\ M_{i}^{2} &=&
 \( 1 + \gamma_i \)  m_{3/2}^2 -
 \wtd{\gamma}_{i} \( \frac{m_{3/2}F^S}{2}+\hc \) \, . \label{BGWsoft2}
\end{eqnarray}
The anomalous dimensions factors $\gamma_{i}$, and the related quantities $\wtd{\gamma}_{i}$, are given in the appendix to this paper in the approximation in which generational mixing can be neglected. 

%%%%%%%%%%%%%%%%%%%%%%%%%%%%%%%%%%%%%%%%%%%%%%%%%%%%%%%%%%%%%%%%%%%%%%
%%%%%%%%%%%%%%%%%%%%%%%%%%%%%%%%%%%%%%%%%%%%%%%%%%%%%%%%%%%%%%%%%%%%%%
\section{BGW Parameter Scan}
\label{scan}

\subsection{Simplification of the Effective Parameter Space}

The preceding section represents an overview of the BGW~model, though even a simplified description of the theory can be quite involved. Nevertheless, the model has relatively few free parameters. This is because at its essence the model is very phenomenological: it is a model of modular-invariant gaugino condensation which invokes the generalized dilaton domination scenario of Casas. Simply put, the BGW model represents {\em any} theory which breaks supersymmetry using gaugino condensation in a hidden sector under the assumptions of the generalized dilaton domination paradigm.

The input parameters can be segregated into various classes which can then be fixed for the subsequent level of analysis. The first step is the minimization of the effective potential~(\ref{Vdil}) for the dilaton field $\ell$. Once a choice such as~(\ref{parameters}) is made for the function~(\ref{ourf}), the phenomenology requires that we satisfy~(\ref{gnew}). While the solution is dependent on the choice of $b_+$, it is only weakly dependent on this choice. Thus we can speak of a `universal solution' for the dilaton potential, leaving us with two quantities,  $\lang \ell \rang$ and $\lang f(\ell) \rang$, relevant for the calculation of the gaugino condensate $u_+$ via~(\ref{uasq}). As mentioned in the previous section, we can use the freedom in the parameterization~(\ref{ourf}) to find a solution such that $\lang f(\ell) \rang \simeq 0$ and thus $\lang \ell \rang \approx g_{\STR}^{2}/2$. With this choice, we are left with the parameter $b_+$ as the sole input parameter at this stage in the calculation.

Under the assumptions described in Section~\ref{modstab} we can describe the gaugino condensate as being determined by (1) the identity of the condensing group with the largest beta-function coefficient, (2) the number of fundamental representations charged under that group, and (3) the effective Yukawa coupling amongst these matter fields in the hidden sector. These three properties determine $b_+$, $(b_+^{\alpha})_{\rm eff}$ and $(c_{\alpha})_{\rm eff}$ (and also $b'_+$ via the identity~(\ref{ba})). The gaugino condensate value is a strong function of these three quantities, and so too is the scale of supersymmetry breaking given by $m_{3/2}$ via~(\ref{mgravplus}).

\begin{figure}[t]
\begin{center}
\includegraphics[scale=0.5]{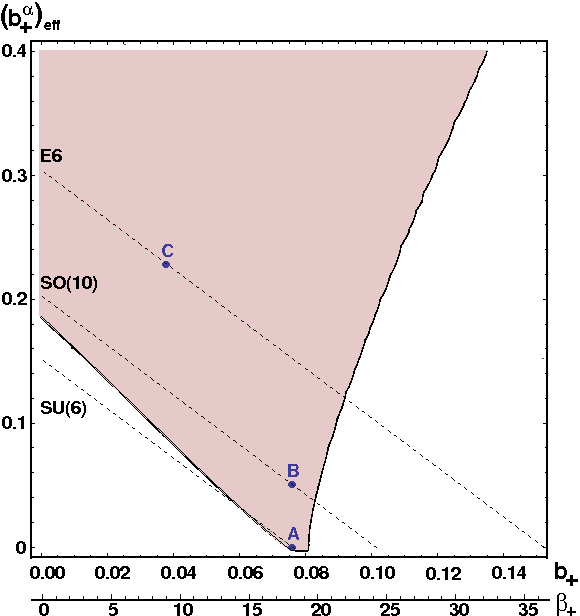}
\caption{\label{plot:gravmass}\textbf{Preferred Region in Group Theory Space.} The shaded region represents the combinations of $b_+$ and $(b_+^{\alpha})_{\rm eff}$ for which the resulting gravitino mass is $m_{3/2} = 10\,{\rm TeV}$. In this plot the value of the effective Yukawa coupling $ (c_{\alpha})_{\rm eff}$ was allowed to vary from $10^{-3}$ (right edge) to $10^3$ (left edge). The labeled points represent example hidden sector configurations: (A) $SU(6)$ with no matter, (B) $SO(10)$ with three \textbf{16} representations, (C) $E_6$ with nine fundamentals. All hidden sector configurations for these three gauge groups will lie on the dotted lines indicated in the plot. The horizontal axis translates the value of $b_+$ into the integer $\beta_+$ used throughout the remainder of the text.}
\end{center}
\end{figure}

However, as was shown in~\cite{Gaillard:1999et}, the dependence of the condensate value on the combination  $(b_+^{\alpha})_{\rm eff}$ and $(c_{\alpha})_{\rm eff}$, via the last product in~(\ref{uasq}), is such that for any value of $b_+ \lappeq 0.12$ one can obtain a realistic gravitino mass of order 1-10~TeV with effective Yukawa couplings in the range $10^{-3} \leq (c_{\alpha})_{\rm eff} \leq 10^3$. This is illustrated in Figure~\ref{plot:gravmass}, where the region in which $m_{3/2} = 10\,{\rm TeV}$ is shown in the $\lbr b_+,\,(b_+^{\alpha})_{\rm eff} \rbr$ plane. We can therefore feel justified, in the BGW~model context, in taking $b_+$ and $m_{3/2}$ as independent parameters provided we restrict ourselves to cases in which $b_+ \lappeq 0.15$.

%=(1)=============== Gravitino mass in group theory space =====================
%==============================================================================

This is a welcome outcome, since the soft supersymmetry breaking terms in~(\ref{gauginomass}) and~(\ref{BGWsoft2}) are manifestly functions of only $m_{3/2}$ and $b_+$, the latter entering through the ratio between $F^S$ and $m_{3/2}$ given by $a_{\rm np}$ in~(\ref{aBGW}). The parameter space that defines the superpartner spectrum is thus effectively two-dimensional. We will therefore begin our investigation taking $b_+$ and $m_{3/2}$ as effectively independent input parameters. In addition, we will follow the common practice of using the known mass of the Z-boson to replace the parameters $\mu$ and $B$ in the electroweak sector with the value of $\tan\beta$. This makes the parameter space effectively three-dimensional.\footnote{We will throughout fix $\mu >0$ as our results are only negligibly affected by the sign of this parameter.} 

Adopting the assumptions that lead to the expression for $b_a$ in~(\ref{ba}) we see that each fundamental charged under the confining group $\mathcal{G}_+$ contributes a fractional amount to the numerator. Our scan results will be far easier to interpret if we therefore re-normalize our expression for the beta-function coefficient to match that of standard supersymmetric renormalization group analyses. We will therefore scan over the quantity 
\begin{equation} \beta_+ = \(3C_+ - \sum_i C_+^i\) \label{betaplus} 
\end{equation}
which we take to be an integer satisfying $3\leq \beta_+ \leq 90$. The upper bound is set by the maximum rank of the confining hidden sector gauge group, which, for the weakly coupled $E_8 \times E_8$ heterotic string would be $C_+ = 30$ for the group $E_8$. The input parameter $b_+$ is then simply
\begin{equation} b_+ = \frac{2}{3} \(\frac{\beta_+}{16\pi^2}\)\, , \label{betaMV} 
\end{equation}
where the quantity in parenthesis is the traditional beta-function coefficient of Martin and Vaughn~\cite{Martin:1993zk}. The correspondence between the two variables is indicated by the two axes in Figure~\ref{plot:gravmass}.

Of course an arbitrary choice of gauge group and matter content need not necessarily exhibit confinement. Nor, for that matter, is such an arbitrary choice guaranteed to be free of anomalies. Strictly speaking, an explicit model of hidden sector gaugino condensation, such as the BGW~model, should restrict the valid choices of $\beta_+$ to reflect these facts. Thus we should not expect the distribution of $\beta_+$ values to be truly flat from the point of view of the underlying string theory. In general, the compactification of weakly-coupled $E_8 \times E_8$ heterotic string theory will favor small values of $\beta_+$, and will struggle to achieve a realistic outcome for $\beta_+ \gappeq 36$.

Furthermore, the variables $\beta_+$ and $m_{3/2}$ are not truly independent, so we should not expect a flat distribution in $m_{3/2}$ when treated as an {\em a priori} variable. 
We will nevertheless do so here in part for the sake of simplicity, but also to better capture the physics of more phenomenological versions of the model, such as the generalized dilaton domination scenario, in which the mechanism of supersymmetry breaking is left unspecified. In such a paradigm it is more legitimate to treat the mass scale $m_{3/2}$ and the parameter $\beta_+$ as independent with flat priors. It is also with such phenomenological scenarios in mind that we will allow $\beta_+$ to be as large as its maximal value of $\beta_+ = 90$, though we will often focus our attention on the smaller values indicated by the explicit BGW~construction. In any event, our goal in this paper is not to present a statistical analysis of the BGW~parameter space, nor make statements as to the most likely region within that parameter space. We merely wish to understand whether the model is still viable after one year of~LHC data and, if so, what are the typical features of the model points that survive.

\subsection{Three-Dimensional Scan}

To fully explore the parameter space, we performed a Monte Carlo scan over the parameters $m_{3/2}$, $\tan\beta$, and $\beta_+$, for which 50,000 random combinations of these parameters was generated, henceforth called `points'. The three parameters were treated as independent variables with flat priors across the ranges $1\,{\rm TeV} \leq m_{3/2} \leq 10\,{\rm TeV}$, $2\leq \tan\beta \leq 50$ and $3\leq \beta_+ \leq 90$. The last quantity was restricted to the domain of integers, as indicated in the previous subsection. 

Having selected a trio of parameters $\lbrace m_{3/2},\,\beta_+,\,\tan\beta\rbrace$, soft supersymmetry-breaking parameters are calculated using~(\ref{gauginomass}) and~(\ref{BGWsoft2}). The ensemble is then evolved to the electroweak scale using {\tt SOFTSUSY 3.3.5}~\cite{Allanach:2001kg} to solve the renormalization group equations. At this stage the radiatively-corrected Higgs potential is minimized and physical masses are calculated.
To restrict our parameter space, we begin with a series of basic requirements on the data set. First, we require that each of these points can achieve electroweak symmetry breaking (EWSB), by which we mean a convergent solution for $\mu$ such that $\mu^2>0$ is found. And we require simultaneously that the lightest superpartner be neutral and colorless. Once this requirement is satisfied, we demand that the superpartner spectrum have sufficiently massive charginos, sneutrinos, and sleptons to have evaded detection via direct searches at~LEP. 

In practice, the only one of these mass limits which constrains the theory is the chargino mass bound, for which we require $m_{\chi_1^{\pm}} > 103.5\,{\rm GeV}$. In the BGW model the parameter $\beta_+$ is bounded from above by the fact that the hidden sector gauge group is presumed to be no larger than $E_8$. This corresponds to a restriction, from~(\ref{aBGW}), of $a_{\rm np} \leq 0.15$, and thus we expect the typical gaugino mass scale to be no larger than a quarter of the scalar masses. Thus, for $\beta_+ =90$ the gravitino mass is bounded from below as $m_{3/2} \geq 1000\,{\rm GeV}$ in order to satisfy the LEP bound on the lightest chargino mass. At smaller values of $\beta_+$, where the gaugino mass suppression is more severe, the lower bound on $m_{3/2}$ is higher. 

%=(2)=============== Wide scan, dark matter constraints ==================
\begin{figure}[t]
\begin{center}
\includegraphics[width=0.75\textwidth]{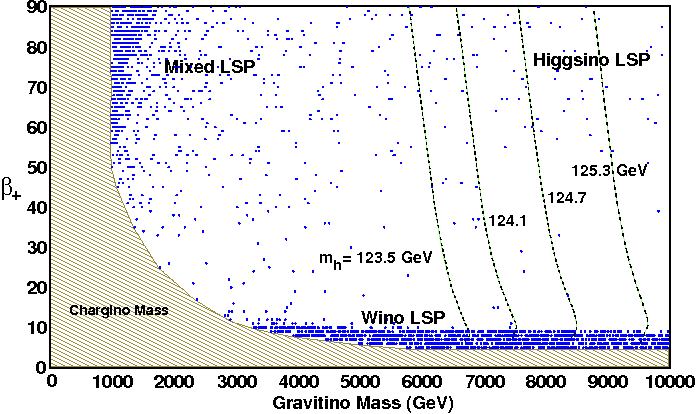}
\caption{\textbf{Dark Matter Constraints.} Distribution of points with proper EWSB and chargino mass, and with a neutralino relic density obeying $\Omega_{\chi}h^2 \leq 0.12$. The shaded region is excluded by the chargino mass constraint $m_{\chi_1^{\pm}} > 103.5\,{\rm GeV}$. The nature of the LSP wavefunction changes dramatically across the plane, as indicated by the labels and described in the text. As a visual reference, contours of constant Higgs mass $m_h$ are given for a fixed value of $\tan\beta=42.5$. The right most contour is the central value of the LHC measurment $m_h = 125.3\,{\rm GeV}$ reported by~CMS, while the contours to the left are the one, two and three sigma lower bounds on the Higgs mass of $m_h = 124.7,\,124.1,\,123.5\,{\rm GeV}$, respectively.}
\label{plot:wideDM}
\end{center}
\end{figure}
%==============================================================================

Given the inherent hierarchy between scalar masses and gaugino masses in the BGW~construction, it is not surprising that the phenomenology of the model will be similar to that of the `focus point'~\cite{Feng:1999zg} or `hyperbolic branch'~\cite{Chan:1997bi} of minimal supergravity. It is well appreciated that results in this region are sensitive to the value of the top mass chosen. In our analysis we have used the central value of the top (pole) mass, $m_t = 173.5\,{\rm GeV}$, as reported by the most recent iteration of the Particle Data Group summary document~\cite{Beringer:1900zz}, which includes the results of recent measurements made at the LHC. Most of the phenomenology which we will discuss below is insensitive to small variations in the top mass. However, the range of $\tan\beta$ values for which a given pair of values $\lbrace m_{3/2},\,\beta_+\rbrace$ can achieve proper EWSB is an important counter-example, as is the mass $m_h$ of the lightest CP-even Higgs eigenstate for a given pair of values $\lbrace m_{3/2},\,\beta_+\rbrace$. We will discuss the top mass sensitivity for these quantities in Section~\ref{finescan} below.

%%%%%%%%%%%%%%%%
Very recently the WMAP~collaboration released their final data analysis, the culmination of nine years of observations~\cite{Bennett:2012fp}. The WMAP data alone is best fit by a relic cold dark matter density of $\Omega_{\chi}h^2 = 0.1138\pm 0.0045$. When added to external data sets representing `extended' CMB measurements, baryon acoustic oscillations, and direct measurements of the Hubble constant, the best fit value becomes  $\Omega_{\chi}h^2 = 0.1153\pm 0.0019$. In interpreting this result for physical models we choose to be as conservative as possible. We therefore allow for the possibility of multi-component dark matter, of which the stable neutralino is but one component, and impose only an upper bound on the neutralino relic density of $\Omega_{\chi}h^2 \leq 0.12$, approximately $2\sigma$ above the mean value of the best fit to the combined data.

Thermal relic abundances of the stable neutralino are computed using {\tt MicrOmegas 2.4.5}~\cite{Belanger:2001fz,Belanger:2004yn}, as are dark matter detection observables discussed later. When the WMAP relic density constraint is imposed the number of points which satisfy the requirement drops to 2,175 of the original 50,000 points. While all values of $\lbrace m_{3/2},\,\beta_+\rbrace$ which pass the EWSB requirements have the {\em potential} to achieve the correct neutralino relic density, the manner in which this is achieved varies across the  $\lbrace m_{3/2},\,\beta_+\rbrace$ plane. The surviving points are displayed in Figure~\ref{plot:wideDM}. The shaded area in the lower left of the figure is already ruled out by the chargino mass bound, arising from direct searches at LEP, as discussed previously. This figures sums over the entire range of scanned $\tan\beta$ values, the distribution of which is mostly flat across the entire scan range. In previous studies of the BGW model~\cite{Gaillard:1999et,Kane:2002qp}, a strong preference for low values of $\tan\beta \lappeq 10$ was found, as dictated by the demands on EWSB. The ability to achieve moderate to large values of $\tan\beta$ in our study is largely the result of improvements made very recently to the treatment of radiative corrections in the Higgs sector with {\tt SOFTSUSY} versions {\tt 3.3.4} and {\tt 3.3.5}~\cite{Allanach:2012qd}.

The correct relic density is most robustly achieved in the region where $5 \leq \beta_+ \leq 10$. In this region the relative values of the wino mass $M_2$ and the bino mass $M_1$ are such that the LSP has a sizable wino fraction. The nature of the cases with $\beta_+ \leq 10$ was studied some time ago in the context of the BGW~model~\cite{BirkedalHansen:2001is,BirkedalHansen:2002am}.  At very small values of $m_{3/2}$ it again becomes relatively easy to satisfy the dark matter constraint. Here the relic density is generally well below the WMAP central value as the LSP is light ($100\,{\rm GeV} \lappeq m_{\chi_1^0} \lappeq 200\,{\rm GeV}$) and admits a large Higgsino content ($|N_{13}|^2 + |N_{14}|^2 \sim 0.5$). Away from these two regions it is still possible to achieve a realistic relic density, but it requires greater tuning of the parameters, particularly the value of $\tan\beta$. In this region, for arbitrary choices of the value of $\tan\beta$, the LSP mass can be quite large and the predicted thermal relic density is as much as three orders of magnitude larger than that required by observations. Here the neutralino is predominantly Higgsino-like ($|N_{13}|^2 + |N_{14}|^2 \sim 1$) with a mass determined largely by the value of $\mu$ which satisfies the EWSB conditions. This value is subject to large radiative corrections -- the same corrections that allow EWSB to occur in the first place -- and are thus subject to theoretical uncertainties associated with how the radiative corrections are implemented into the computer code~\cite{Allanach:2012qd}.

%=(3)===== Wide scan, dark matter constraints, beta+ histogram ===========
\begin{figure}[t]
\begin{center}
\includegraphics[width=0.7\textwidth]{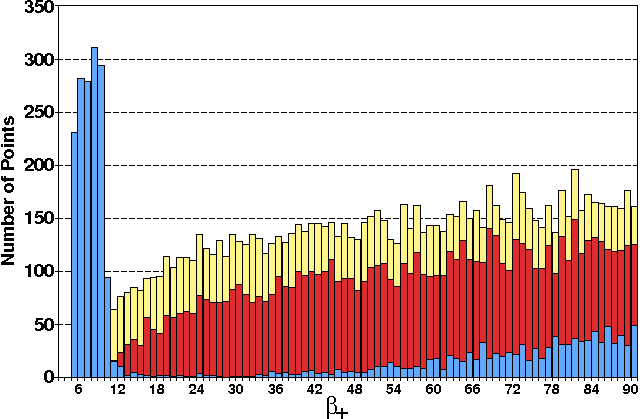}
\caption{\textbf{Histogram of Allowed $\beta_+$ Values, Dark Matter Preferred Regions versus Higgs Mass.} The blue bars represent the distribution in $\beta_+$ values with the dark matter constraint $\Omega h^2 <0.12$ imposed, but no Higgs mass constraint. The red bars are $\beta_+$ values with $m_h>124.1\,{\rm GeV}$, and no dark matter constraint. The yellow bars are $\beta_+$ values with $m_h>123.5\,{\rm GeV}$, and no dark matter constraint.}
\label{plot:DMhisto}
\end{center}
\end{figure}
%==============================================================================

Data analyzed in~2012 revealed evidence of a Higgs-like boson at both the ATLAS and CMS~experiments. In the case of CMS~\cite{:2012gu} the early data indicated a mass of $m_h = 125.3 \pm 0.6\,{\rm GeV}$, where we have added the reported statistical and systematic errors in quadrature. For ATLAS~\cite{:2012gk} the central value was slightly higher, with $m_h = 126.0 \pm 0.6\,{\rm GeV}$, again adding the reported errors in quadrature. In recent days these measurement have been updated with additional data, and a mass closer to 126~GeV seems to be preferred. Nevertheless, for this analysis we will be conservative and allow for a three-standard deviation range about the~CMS value of $m_h = 125.3$ as our constraint on the Higgs mass. 
In practice, for the BGW~model this only enforces a lower bound, that is $m_h \geq 123.5\,{\rm GeV}$. For the three-dimensional scan of 50,000 generated points, 11,189 satisfied this requirement on the mass of the lightest CP-even Higgs. These points all satisfy $\beta_+ \geq 10$ and $m_{3/2} \geq 6.9\,{\rm TeV}$. The contours for the central value and the one-, two- and three-sigma lower bounds on the Higgs mass $m_h$ are shown in the upper right of Figure~\ref{plot:wideDM}.

%=(4)==== Wide scan, dark matter constraints, Higgs mass histogram =========
\begin{figure}[t]
\begin{center}
\includegraphics[width=0.7\textwidth]{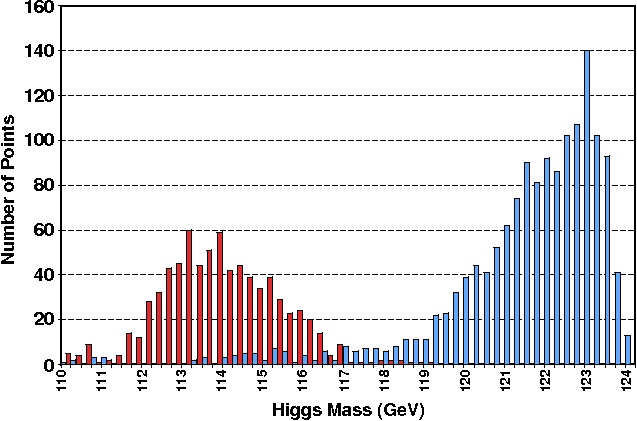}
\caption{\textbf{Histogram of Higgs Masses, Dark Matter Preferred Regions.} The distribution in $m_h$ values with the dark matter constraint $\Omega h^2 <0.12$ imposed, for all values of $\tan\beta$. The wino-like region ($\beta_+ \leq 10$) is described by the blue bars. The mixed LSP case with low $m_{3/2}$ is described by the red bars. As described in Section~\ref{finescan}, achieving $m_h \simeq 126\,{\rm GeV}$ will require fixing $\tan\beta \gappeq 42$.}
\label{plot:Higgshisto}
\end{center}
\end{figure}
%==============================================================================

The Higgs mass constraint from the recent LHC~measurements is in some degree of tension with the results from the WMAP~satellite for the BGW~model. The distribution of $\beta_+$ values across the 2,175~points with a relic density $\Omega_{\chi}h^2 \leq 0.12$ is given by the blue histogram in Figure~\ref{plot:DMhisto}. Meanwhile, the distribution of the 11,189~points with $m_h \geq 124.1\,{\rm GeV}$ and $m_h \geq 123.5\,{\rm GeV}$ is given by the red and yellow histograms, respectively, in the same figure. There is a small overlap at $\beta_+ \simeq 10$ if a two-sigma lower bound on the Higgs mass is employed, though most of the wino-like region is viable if a looser three-sigma lower bound is utilized. 

The distribution in calculated $m_h$ values for the lightest CP-even Higgs is given in Figure~\ref{plot:Higgshisto} for the two areas of the BGW~parameter space for which the dark matter relic density constraint is readily satisfied. The area with large $\beta_+$ but low $m_{3/2}$ tends to favor the region $112\,{\rm GeV} \lappeq m_h \lappeq 117\,{\rm GeV}$. These Higgs masses were already being constrained by the searches at~LEP. For the promising wino-like region the tension is less: Higgs masses are generally in the window $120\,{\rm GeV} \lappeq m_h \lappeq 124\,{\rm GeV}$, just below the two-sigma lower bound on the Higgs mass using the~2012 LHC~results. Away from these two areas the Higgs mass measurement and neutralino relic density constraint can still be reconciled, provided the value of $\tan\beta$ is reasonably large. With the value of $m_t = 173.5\,{\rm GeV}$, a sufficiently heavy CP-even Higgs mass arises once $\tan\beta \geq 42.5$. It is important to note, however, that the region of the parameter space for which a Higgs mass $m_h \geq 123.5\,{\rm GeV}$ can be achieved is sensitive to the value of the input top quark mass. For higher values of $m_t$, more of the wino-like region at low $\beta_+$ values becomes allowed. 

\subsection{Two-Dimensional Scan}
\label{finescan}

%=(5)=============== Fine scan, EWSB only ==================
\begin{figure}[t]
\begin{center}
\includegraphics[width=1.00\textwidth]{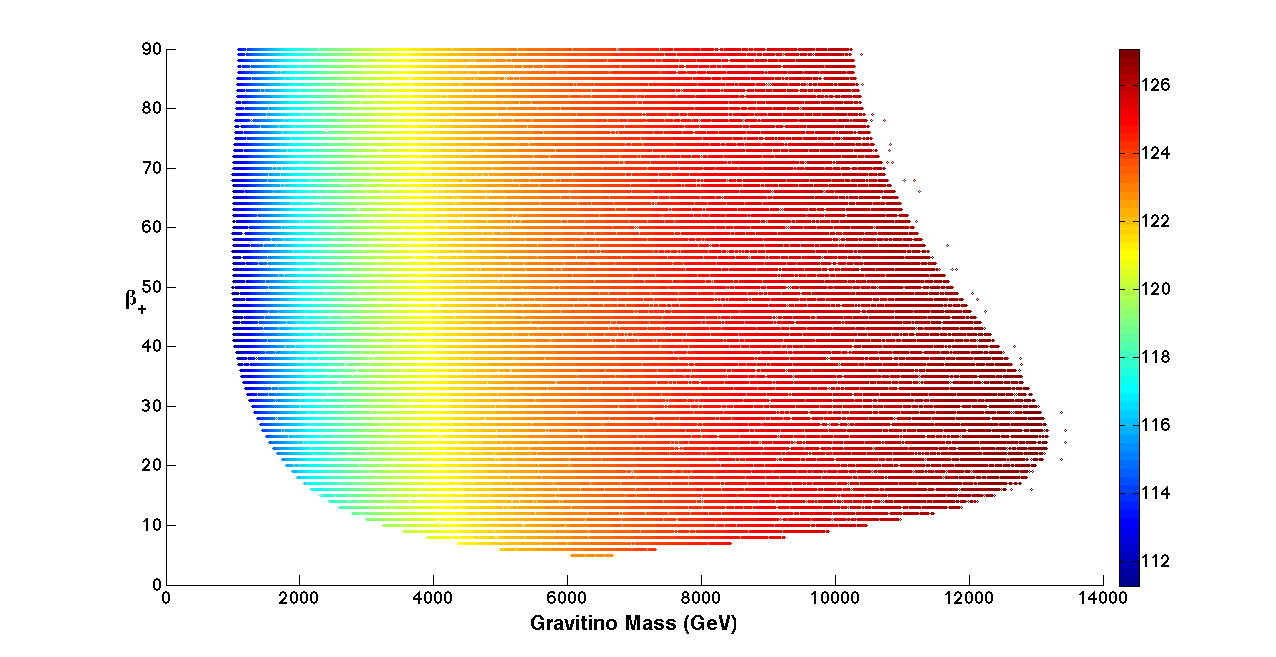}
\caption{\textbf{Two-Dimensional Scan of BGW Parameter Space for $\tan\beta=42.5$.} Distribution of points with proper EWSB and chargino mass, for $m_t = 173.5\,{\rm GeV}$. The Higgs mass (in units of GeV) is indicated by the color. The region in the lower left is excluded by the chargino mass bound as before. The region to the right at extreme values of $m_{3/2}$ is excluded by a lack of reliable EWSB.}
\label{plot:fineall}
\end{center}
\end{figure}
%==============================================================================

In order to generate enough data to be able to adequately visualize the phenomenology of the remaining parameter space, we perform a second scan using the same methodology as before, but in this instance we will fix the value of $\tan\beta=42.5$. Again 311,111 points are generated across the $\lbrace m_{3/2}, \beta_+ \rbrace$ plane, this time with $1\,{\rm TeV} \leq m_{3/2} \leq 15\,{\rm TeV}$ and $3\leq \beta_+ \leq 90$. The points which passed the initial requirements of proper EWSB and sufficiently heavy chargino and sleptons are shown in Figure~\ref{plot:fineall}.

As before, the area in the lower left of the figure is excluded by the requirement that the chargino mass satisfy $m_{\chi_1^{\pm}} > 103.5\,{\rm GeV}$. Beyond $m_{3/2} = 10\,{\rm TeV}$ the radiative corrections to the Higgs potential eventually become large enough to drive the effective value of $\mu$ to below about 200~GeV. At this point, the convergence on the calculated value of $\mu^2$ which satisfies the EWSB requirement becomes quite poor. The nature of this region was the focus of a detailed study~\cite{Allanach:2012qd}, using the most recent version of {\tt SoftSUSY}. We have decided to take the onset of this poor convergence region as an upper bound on the gravitino mass we will consider. In some sense this is a theoretical prejudice: the question of whether the parameter space beyond this region is truly valid must await further theoretical improvements in the treatment of radiative corrections to the Higgs potential. Operationally speaking, the lack of convergence on the calculation of $\mu$ will imply that phenomenology involving the lightest neutralino will become increasingly suspect in this region, so we have chosen to be prudent by only displaying results for which {\tt SoftSUSY 3.3.5} returns a result with $\mu \geq 300\,{\rm GeV}$.

We are forced to explore this extreme region of the parameter space by the Higgs mass measurement at the LHC. The color in Figure~\ref{plot:fineall} indicates the mass $m_h$ of the lightest CP-even Higgs eigenstate. The demand that $m_h \geq 123.5\,{\rm GeV}$ requires $m_{3/2} \gappeq 6.2\,{\rm TeV}$, as was already apparent from the three-dimensional scan results. The LHC~upper bound, at the two sigma level, already includes most of the parameter space up to the region of poor convergence. The ability to achieve a Higgs mass $m_h \simeq 126\,{\rm GeV}$, while simultaneously achieving reliable EWSB, necessitates the choice of $m_t = 173.5\,{\rm GeV}$. For example, using the best-fit value reported by the~PDG in~2011, $m_t = 172.9\,{\rm GeV}$, would have placed the bulk of the parameter space in which $m_h \geq 125.3$ beyond the region where EWSB becomes unreliable.

%=(6)=============== Fine scan, Higgs constraints, DM in color ==================
\begin{figure}[t]
\begin{center}
\includegraphics[width=1.00\textwidth]{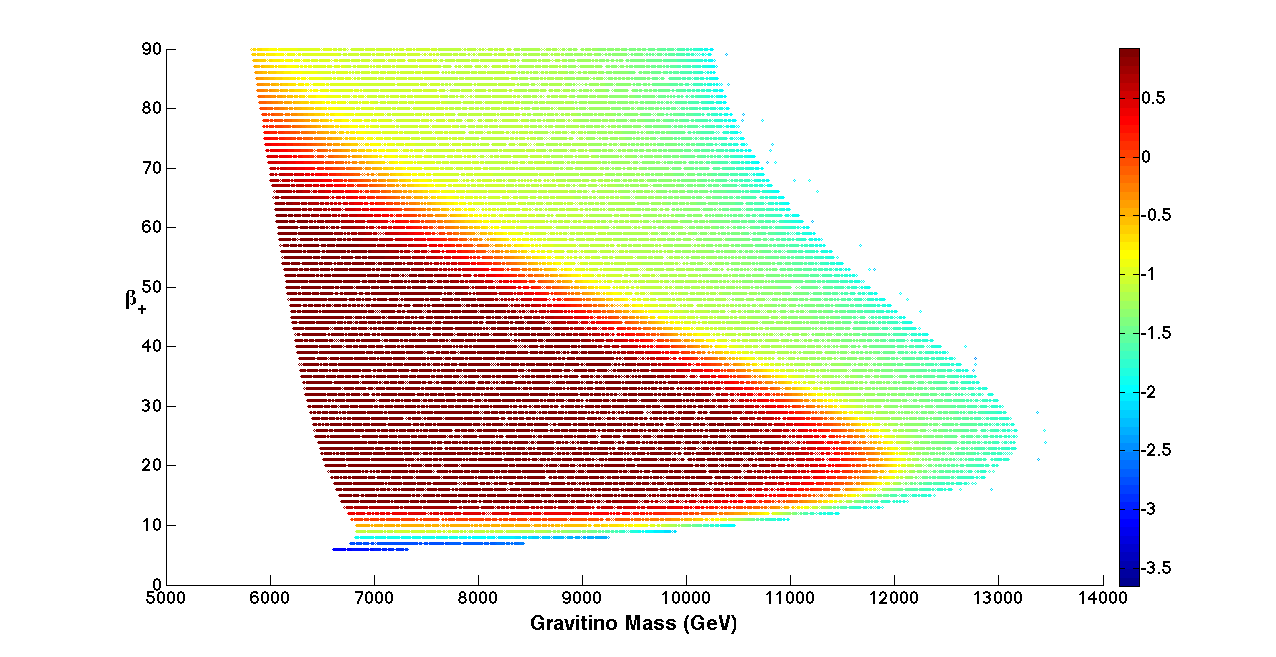}
\caption{\textbf{Neutralino Relic Density for $\tan\beta=42.5$ and $m_h \geq 123.5\,{\rm GeV}$.} Distribution of points with proper EWSB, chargino mass, and Higgs mass. The empty region to the left has $m_h <123.5\,{\rm GeV}$. The empty region to the right is excluded by a lack of reliable EWSB. The neutralino relic density is calculated on a logarithmic scale $\log_{10}\(\Omega h^2\)$ and is indicated by the color. The interior red-colored points with large relic densities correspond to points with an overwhelmingly bino-like LSP.}
\label{plot:fineDM}
\end{center}
\end{figure}
%==============================================================================

%=(7)=============== Fine scan, all constraints, gluino in color ==================
\begin{figure}[th]
\begin{center}
\includegraphics[width=1.00\textwidth]{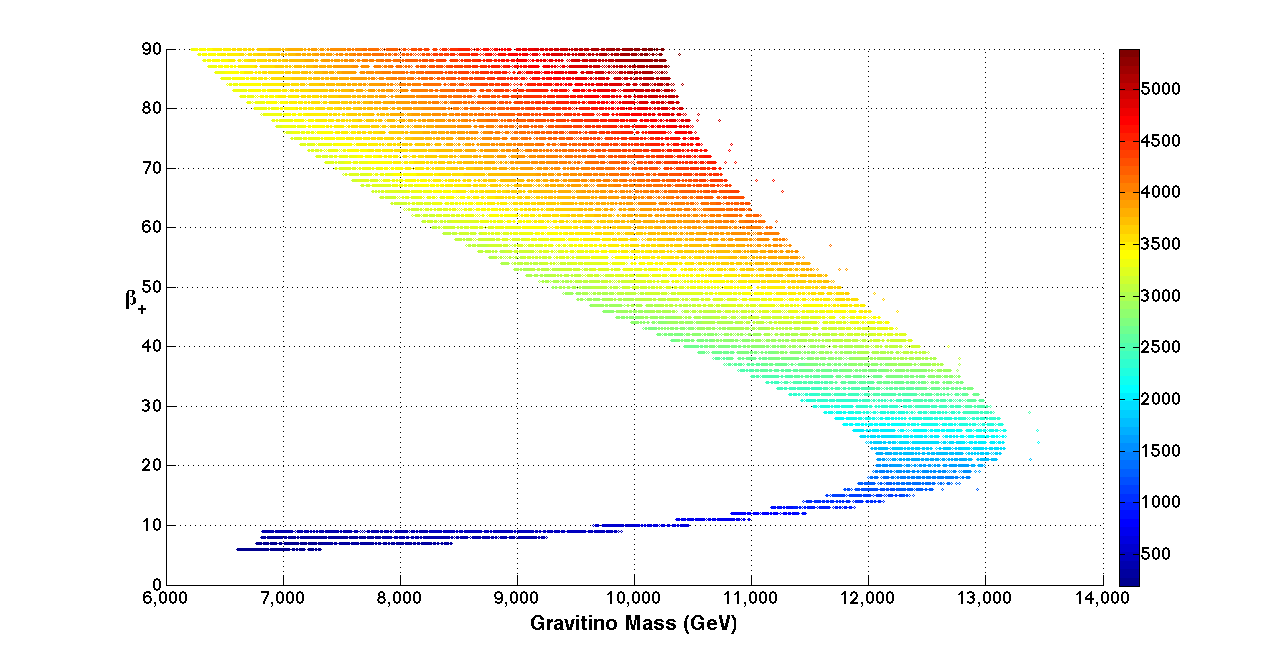}
\caption{\textbf{Final Allowed BGW Parameter Space for $\tan\beta=42.5$.} Distribution of points with proper EWSB, chargino mass, Higgs mass, and neutralino relic density. The empty region to the left has $m_h <123.5\,{\rm GeV}$ and/or $\Omega h^2 > 0.12$. The empty region to the right is excluded by a lack of reliable EWSB. The gluino mass (in units of GeV) is indicated by the color.}
\label{plot:fineGluino}
\end{center}
\end{figure}
%==============================================================================

Having therefore imposed the lower bound $m_h \geq 123.5\,{\rm GeV}$ for $m_t = 173.5\,{\rm GeV}$, we indicate the calculated thermal relic density for the remaining points in Figure~\ref{plot:fineDM}. The horizontal scale now begins at $m_{3/2} = 5\,{\rm TeV}$, and the color of the point indicates the neutralino relic density on a logarithmic scale $\log_{10}\(\Omega h^2\)$. The WMAP preference is then for points with a value $\log_{10}\(\Omega h^2\) \lappeq -1$, which is roughly the yellow band in Figure~\ref{plot:fineDM}. A great deal of the parameter space is immediately eliminated for having a thermal dark matter abundance far too high relative to the WMAP measurement. At the lowest values of $\beta_+$ the LSP is overwhelmingly wino-like and thus $\Omega h^2$ is quite small. As the value of $\beta_+$ increases, for a fixed mass scale $m_{3/2}$, the LSP rapidly becomes entirely bino-like and the annihilation rate for the lightest neutralino in the early universe drops precipitously. This persists until one reaches the region in which radiative corrections come to dominate. Here the LSP begins to develop a sizeable Higgsino component as the radiatively-corrected value of $\mu$ diminishes. This behavior, and the general area of the parameter space eliminated by WMAP constraints, is not measureably affected by changes in the top mass.

The parameter space for the BGW~model, in its simplest manifestation, is therefore constrained from all sides, though the upper bound on the mass scale $m_{3/2}$ is a result of theoretical uncertainty in handling radiative corrections to EWSB. The final allowed parameter space is shown in Figure~\ref{plot:fineGluino}, which includes only those points with $m_{\chi_1^{\pm}} > 103.5\,{\rm GeV}$, $m_h \geq 123.5$, $\Omega h^2 \leq 0.12$ and sufficiently convergent EWSB. The right-most edge of this region is mildly sensitive to the value of $m_t$. The left-most edge is sensitive to the value of $\tan\beta$ chosen. In Figure~\ref{plot:fineGluino} these values are $m_t=173.5\,{\rm GeV}$ and $\tan\beta= 42.5$, respectively.

Having an effectively finite parameter space allows us to make definite predictions for the model. For example, this model is consistent with all indirect searches for supersymmetry. For example, the branching ratio $B_s\rightarrow\mu^+\mu^-$ takes a value $3.05\pm0.02\times10^{-9}$ across this space, in perfect agreement with the experimental results and Standard Model predictions. This is largely the result of the very heavy scalars in the allowed parameter space. Similarly, the model is perfectly consistent with the Standard Model for other rare decays, such as $b \to s \gamma$ decays. The remaining parameter space also predicts a vanishingly small contribution to the anomalous magnetic moment of the muon, with $8.0\times 10^{-12} \leq \delta a_{\mu} \leq 4.6 \times 10^{-11}$. 
It continues to be an open question as to how consistent experimental measurements of $g_{\mu}-2$ are with the theoretical predictions of the Standard Model~\cite{Melnikov:2003xd,Jegerlehner:2009ry}. Clearly, the BGW model does not imply a new physics contribution to this quantity.

Despite the large mass scale, superpartners are nevertheless within reach of the LHC, particularly after the shutdown and upgrade to higher center-of-mass energies. Points in Figure~\ref{plot:fineGluino} are colored according to the predicted gluino mass for that point. The allowed parameter space predicts gluinos as light as 200~GeV, reaching a maximum mass, for $\beta+ = 90$, of 5400~GeV. Clearly, gluino masses on the low end of this range are excluded by the LHC search limits in channels such as multijets and missing transverse energy. We will discuss these limits in the next section. Here we wish merely to point out that the string theory preferred region of $\beta_+ \leq 36$ puts an upper bound on the gluino mass of $m_{\tilde{g}}\leq 2910\,{\rm GeV}$; requiring $\beta_+ \leq 24$ (as implied by Figure~\ref{plot:gravmass}) would predict $m_{\tilde{g}}\leq 2040\,{\rm GeV}$. Meanwhile the gluino mass exceeds 1~TeV only for $\beta_+ \geq 14$. Thus, the theoretically interesting region of the parameter space is already being probed with current LHC~data, and most of this interesting region is within reach at $\sqrt{s}=13\,{\rm TeV}$.

%%%%%%%%%%%%%%%%%%%%%%%%%%%%%%%%%%%%%%%%%%%%%%%%%%%%%%%%%%%%%%%%%%%%%%%%%%%%%%%%
%%%%%%%%%%%%%%%%%%%%%%%%%%%%%%%%%%%%%%%%%%%%%%%%%%%%%%%%%%%%%%%%%%%%%%%%%%%%%%%%
\section{LHC Implications}
\label{LHC}

\subsection{Discussion of Benchmark Points}

In the previous section, a region of parameter space for the BGW model was established which is in agreement with the WMAP constraint on the density of neutralino dark matter, as well as the combined ATLAS/CMS discovery of a Higgs boson with mass of approximately 125-126~GeV. We also established that the masses of superpartners were sufficiently heavy so as to avoid direct discovery at pre-LHC collider experiments. Specifically, however, we did not impose a strict lower bound on the gluino mass, which is constrained not only by search limits from the~LHC, but also by Tevatron search results. As we will see shortly, the gluino is within the current experimental reach of ATLAS and CMS for low values of $\beta_+$, and should be accessible after the shutdown for much of the theoretically interesting parameter space.

To determine what part of the parameter space is already constrained by searches for supersymmetry at the~LHC, and what area is within reach in the near future, we should analyze simulated collider data at each point in the remaining parameter space, using the techniques employed by the two general-purpose detectors at the~LHC. This is computationally prohibitive, but we are fortunate in that the parameter space of the BGW model can be made effectively two-dimensional, as was seen in Section~\ref{finescan}. Furthermore, Figure~\ref{plot:fineGluino} indicates that the mass of the gluino is roughly proportional to $\beta_+$ (and independent of the gravitino mass) for much of the parameter space with $\beta_+ \leq 36$. This makes the model effectively one-dimensional. We can therefore feel confident that it is sufficient to sample the parameter space via a few well-chosen benchmark points and interpolate between them to get a general feel for the LHC~implications of the BGW~model.

%%%%%%%%%%%%%%%%%% BENCHMARK TABLE %%%%%%%%%%%%%%%%%%%%%%%%%
\begin{table}[t]
\begin{center}
\begin{tabular}{|c||c|c||c|c|c|c|} \hline
 &  \multicolumn{2}{|c||}{BGW Parameters} & \multicolumn{4}{|c|}{Key Physical Masses (GeV)} \\ \hline
Point & \parbox{1.0cm}{$\beta_+$} & \parbox{1.0cm}{$m_{3/2}$} &\parbox{0.9cm}{$m_{\tilde{g}}$} & \parbox{0.9cm}{$m_{\tilde{N}_1}$} & \parbox{0.9cm}{$m_{\tilde{C}_1}$} & \parbox{0.9cm}{$m_h$} \\ \hline
% 51985
A & 9 & 9425 & 498 & 246 & 263 & 125.3 \\ 
% 227445
B & 10 & 10465 & 628 & 266 & 283 & 125.9 \\
% 217701
C & 11 & 10360 & 699 & 309 & 345 & 125.7 \\
% 218646
D & 12 & 10828 & 808 & 340 & 381 & 126.0 \\
% 231808
E & 13 & 11175 & 913 & 369 & 413 & 126.1 \\
% 225426
F & 14 & 11870 & 1050 & 366 & 383 & 126.4 \\
% 218022
G & 15 & 11649 & 1114 & 422 & 467 & 126.3 \\
% 212095
H & 18 & 11994 & 1392 & 492 & 535 & 126.4 \\
% 222261
I & 24 & 11983 & 1866 & 607 & 645 & 126.5 \\ 
% 232164
J & 33 & 11457 & 2437 & 707 & 723 & 126.3 \\ 
% 223734
K & 42 & 11369 & 3031 & 637 & 642 & 126.3 \\ 
% 106272
L & 60 & 8279 & 3099 & 869  & 891 & 124.9 \\ 
% 233646
M & 60 & 10502 & 3877 & 577 & 581 & 126.0 \\ 
% 4253
N & 81 & 9996 & 4800 & 498 & 500 & 125.9 \\ 
\hline
\end{tabular}
\caption{\textbf{Benchmark Points for Phenomenological Analysis}. Sample points from the BGW parameter space which pass all pre-LHC phenomenological constraints. All points have $\mu>0$ and $\tan\beta=42.5$. Masses for the gravitino, gluino, lightest neutralino $\tilde{N}_1$, lightest chargino $\tilde{C}_1$ and the lightest Higgs eigenstate are given in~GeV. Physical squark and slepton masses are roughly 10~TeV for all benchmarks.}
\label{masstable}
\end{center}
\end{table}
%%%%%%%%%%%%%%%%%% BENCHMARK TABLE %%%%%%%%%%%%%%%%%%%%%%%%%

In order to discuss the reach of the LHC~experiments in the $\lbrace \beta_+,\,m_{3/2} \rbrace$ plane -- and to illustrate correlations between LHC phenomenology and future dark matter direct detection experiments -- we have chosen a set of 14~benchmark points from Figure~\ref{plot:fineGluino} which span the range $9 \leq \beta_+ \leq 81$. These benchmarks are shown in Table~\ref{masstable}, where the physical masses of the gluino, lightest neutralino $\tilde{N}_1$, lightest chargino $\tilde{C}_1$ and the lightest Higgs eigenstate are given in~GeV. These points cover nearly an order of magnitude in gluino mass (from 500~to 5000~GeV), though the squark and slepton masses are roughly 10~TeV for all benchmark points. Note that the average Higgs mass across these 14~benchmarks is 126.0~GeV.

For each benchmark point, we compute the high scale boundary terms using~(\ref{gauginomass}) and~(\ref{BGWsoft2}). We then evolve these values to the electroweak scale using {\tt SOFTSUSY 3.3.5}, which also generates the physical masses and couplings of the superpartners. Calculation of decay widths and branching ratios is then performed using {\tt SUSY-HIT}~\cite{Djouadi:2006bz}. This information is then passed to {\tt PYTHIA 6.4}~\cite{Sjostrand:2006za} for event generation and {\tt PGS4}~\cite{PGS4} to simulate the detector response. For each point we consider, we generate a fixed number of 50,000~events in order to achieve accurate results which can then be scaled to the integrated luminosity appropriate to any particular published result. In practice, for all of our benchmarks, this number of events represents an integrated luminosity well above the amount thus far analyzed.

In what follows, the analysis will necessarily be at a somewhat superficial level. For example, we use the general purpose {\tt PGS4} software to simulate detector response without imposing any trigger requirements on the signal. In what follows we will estimate the reduction of the signal when triggering efficiencies are considered by imposing the level-one trigger requirement contained within {\tt PGS4}, but all results will always be displayed without triggers. In addition, we will consider only the searches conducted by the ATLAS detector. We do so for two reasons. First, the particle identification requirements and methodology for computing such objects as missing transverse energy ($E_{T}^{\rm miss}$) are nearly universal across all ATLAS SUSY searches. Second, ATLAS search strategies have thus far been built from simple geometric cuts on individual detector objects, and certain simple kinematic variables constructed from these objects. The CMS collaboration, by contrast, often utilizes more sophisticated objects such as $\alpha_T$~\cite{Randall:2008rw}, $M_{T2}$~\cite{Lester:1999tx} and the razor distribution~\cite{Rogan:2010kb}. Thus, the CMS searches are often designed in such a way as to make use of a complete knowledge of detector geometry and calorimetry that is not available to the theorist. The overall reach for superpartners at CMS is not substantially different from that obtained at ATLAS, however. Finally, we will simulate only the signal for our benchmark points, and not the underlying Standard Model backgrounds. We use the reported event rates and estimated cross-section limits directly from the ATLAS searches to determine compatability of a given benchmark with the data. In this paper our goal is to understand how LHC data is beginning to impact {\em bona fide} models of string theory, starting with one of the most simple models available. We therefore feel that this level of analysis is suitable, particularly given the sizeable theoretical uncertainties described in the previous sections.

%-------------------------------------------------------------
\subsection{ATLAS Searches and Signal Regions}
\label{searches}

ATLAS has published the results of over a dozen SUSY searches for data taken at 8~TeV, allowing several possible avenues for discovery. These results represent integrated luminosities between 5.8~and 20.7~fb$^{-1}$, with the more generic searches having been performed earlier and thus typically involving smaller data sets.\footnote{We will always scale our simulated data to the appropriate integrated luminosity, using the total SUSY production cross-section as reported by {\tt PYTHIA} as our guide.} The more recent searches have been optimized for special circumstances in the parameter space of supersymmetric models, such as preferential production of third-generation squarks, R-parity violation, and gauge-mediated models. These cases tend to emphasize high-p$_T$ leptons and b-tagged jets in the event selection. Many of these dedicated searches are not relevant for the BGW~model. For example, a typical event for any of the benchmarks in Table~\ref{masstable} has zero or one high-p$_T$ electron or muon; cases with two or more such leptons are quite rare. In addition, mass differences between light electroweak gauginos are typically too small to generate any on-shell Z-bosons in the gaugino decays.
On the other hand, these benchmark points tend to generate a fairly large number of b-tagged jets.
We will therefore focus our efforts on jet-based searches involving large missing transverse energy, few leptons and b-tagging. To that end, we use the same modifications to the internal b-tagging algorithm of {\tt PGS4} as was described in~\cite{Altunkaynak:2010we}, which better matches the estimated b-tagging efficiency of the ATLAS~detector.

Each of these searches differs slightly in object reconstruction, but for the most part a standard set of object identification requirements is imposed across all searches. These criteria are as follows:
\begin{itemize}
\item Jets are typically required to have a minimum transverse momentum $p_T > 20\,{\rm GeV}$ and $|\eta|<2.8$. %Jets are reconstructed using the anti-$k_T$ algorithm with a distance radius of~0.4.
\item Electrons are required to have $p_T>20\,{\rm GeV}$ and $|\eta|<2.47$ while muons must have $p_T>10 \,{\rm GeV}$ and $|\eta|<2.4$.
\item An isolation criterion is applied wherein if $\Delta R\equiv\sqrt{(\Delta\eta)^2 + (\Delta\phi)^2}<0.2$ between an electron and any given jet, the jet candidate is discarded, and any lepton within $\Delta R = 0.4$ of a jet is discarded.
\item The missing transverse momentum is calculated as the negative vector sum of the x- and y- components of the reconstructed transverse momenta of all surviving jets and leptons.
\end{itemize}
The above object definitions are imposed universally across all analyses we perform. Note that the last item in the list requires a modification to the default calculation of missing transverse energy performed within {\tt PGS4}.

We briefly summarize the defining characteristics of each of these searches and their signal regions below. In what follows, we will follow the ATLAS collaboration in making use of several variants of the effective mass quantity $m_{\rm eff}$. Most often, this variable represents the scalar sum of the transverse momenta of the leading $N_j$ jets, together with the missing transverse momentum. In these cases it is denoted $m_{\rm eff}(N_j)$. For the low- and high-multiplicity jet searches the inclusive effective mass $m_{\rm eff}({\rm inc.})$ is simply the scalar sum of the transverse momenta of all jets with $p_T > 40\,{\rm GeV}$. In the case of the high-multiplicity jets search this variable is denoted $H_T$.  For the single lepton analysis, two different effective masses were used. One was the inclusive effective mass, which included all jets with a $p_T$ above 40~GeV, and the single hardest lepton. The other was simply the effective mass coming from the four hardest jets and the one hardest lepton. For further details, the reader is encouraged to visit the referenced conference notes~\cite{twiki}.

%%%%%%%%%%%%%%%%%%%%%%%%%%%%%%%%%%%%%%%%%%%%%
%ATLAS-CONF-2012-103, ATLAS-CONF-2012-104, ATLAS-CONF-2012-105, ATLAS-CONF-2012-109, ATLAS-CONF-2012-147, and ATLAS-CONF-2012-152. 

\begin{description}
\item[Low Jets (ATLAS-CONF-2012-109: $\mathbf{\mathcal{L} = 5.8\,{\rm fb^{-1}}}$)] This search contains 12 distinct signal regions. For each, there must be no reconstructed leptons, a minimum number of between 2~and 6~jets with the leading jet having $p_T>160\,{\rm GeV}$ and subsequent jets having $p_T>60\,{\rm GeV}$. The three hardest jets (when applicable) must be separated from the missing transverse energy by  $\Delta\phi>0.4$; any applicable subsequent jets must have only $\Delta\phi>0.2$. All channels require $E_T^{\rm miss} > 160 \, {\rm GeV}$. A requirement is placed on the inclusive effective mass varying for each signal region but ranging between 1000~GeV and 1900~GeV. Finally, the ratio $E_T^{\rm miss}/m_{\rm eff}(N_j)$ is restricted to be a minimum of between 0.15 and 0.4, depending on the signal region.
\item[High Jets (ATLAS-CONF-2012-103: $\mathbf{\mathcal{L} = 5.8\,{\rm fb^{-1}}}$)] This search contains 6~signal regions. In addition to a veto on reconstructed leptons, a minimum jet multiplicity requirement of between 6~and 9~jets is imposed. Depending on the signal region, these jets must have have a minimum $p_T$ of either 55~or 80~GeV per jet. Finally, while there is no requirement on the absolute size of the missing transverse energy, a requirement is placed on $E_T^{\rm miss}/\sqrt{H_T}>4\,{\rm GeV}^{1/2}$. 
\item[Single Lepton (ATLAS-CONF-2012-104: $\mathbf{\mathcal{L} = 5.8\,{\rm fb^{-1}}}$)] The single lepton search defines two signal regions: one with a single electron, and one with a single muon. For each, $p_T^{\ell}>25\,{\rm GeV}$, with no other reconstructed leptons. There must be at least four jets with $p_T>80\,{\rm GeV}$.  Furthermore, each signal region requires $E_T^{\rm miss}>250\,{\rm GeV}$, $m_{\rm eff}>800\,{\rm GeV} $,  $E_T^{\rm miss}/m_{\rm eff}>0.2$, and $m_T>100{\rm GeV}$, where $m_T$ is the standard transverse mass variable formed from the single lepton and the missing transverse momentum. 
\item[Same Sign Dilepton (ATLAS-CONF-2012-105: $\mathbf{\mathcal{L} = 5.8\,{\rm fb^{-1}}}$)] This search requires at least two leptons ($e$ or $\mu$) with the same sign and $p_T > 20\,{\rm GeV}$. Three signal regions: $e\,e$, $e\,\mu$, and $\mu\,\mu$ are then defined, as well as a combined result for $\ell\ell$. A minimum of 4~jets are required with $p_T > 50\,{\rm GeV}$, as well as a minimum $E_T^{\rm miss}>150\,{\rm GeV}$. 
\item[SS Dilepton + B-Jets (ATLAS-CONF-2013-007: $\mathbf{\mathcal{L} = 20.7\,{\rm fb^{-1}}}$)] For this search there must be at least two leptons ($e$ or $\mu$) with the same sign and $p_T > 20\,{\rm GeV}$. Three signal regions are then defined. The first (SR 0b) imposes a veto on b-tagged jets, requires at least~3 non-b-jets with $p_T$ threshold of $p_T>40\,{\rm GeV}$ and $E_T^{\rm mis}>150\,{\rm GeV}$. In addition the transverse mass formed from the hardest lepton and the missing transverse momentum must satisfy $m_T>100\,{\rm GeV}$, and the effective mass formed from the two hardest leptons, jets and missing transverse energy must be greater than 400~GeV. A second signal region (SR 1b) requires at least 1 b-tagged jet, and increases the effective mass requirement to 700~GeV. The third signal region requires only that there be at least 4~jets and a minimum of 3~b-tagged jets. For this analysis, the reconstruction criteria are altered from the previous analysis: the minimum $p_T$ for a b-tagged jet remains 20~GeV, while non-b-tagged jets require a minimum $p_T$ of 40~GeV. The minimum muon $p_T$ is raised to 20~GeV.
\end{description}

%-------------------------------------- ATLAS SR RESULTS ---------------------------
\begin{table}[th]
\begin{center}
\begin{tabular}{|l|c|c|c||l|c|c|c|}
\hline
Signal & Observed & \multicolumn{2}{|c||}{95\% Upper Limit}  & Signal & Observed &   \multicolumn{2}{|c|}{95\% Upper Limit}  \\
Region & Events &  $\sigma_{\rm BSM}$ (fb) & \parbox{1.0cm}{$N_{\rm BSM}$} & Region & Events &  $\sigma_{\rm BSM}$ (fb) & \parbox{1.0cm}{$N_{\rm BSM}$} \\ \hline\hline
2~jets (loose)  & 643 & 38.8 & 225		& 7~jets (55~GeV) & 381 & 21   & 122    \\ \hline
2~jets (medium) & 111 & 5.8  & 34		& 8~jets (55~GeV) & 48  & 5    & 29 \\ \hline
2~jets (tight)  & 10  & 1.5  & 9		& 9~jets (55~GeV) & 3   & 0.9  & 5 \\ \hline
3~jets (medium) & 106 & 7.6  & 44		& 6~jets (80~GeV) & 248 & 15.7 & 91 \\ \hline
3~jets (tight)  & 7   & 1.3  & 7		& 7~jets (80~GeV) & 26  & 4.3  & 25 \\ \hline
4~jets (loose)  & 156 & 11.3 & 66		& 8~jets (80~GeV) & 1   & 0.7  & 4 \\ \hline
4~jets (medium) & 31  & 3.1  & 18		& 1 lepton ($e$)  & 10  & 1.7  & 10 \\ \hline
4~jets (tight)  & 1   & 0.6  & 3		& 1 lepton ($\mu$)& 4   & 1.1  & 6 \\ \hline
5~jets          & 5   & 1.0  & 6		& SS 2$\ell$ + 0~~~~B-jets     & 5 & 0.3 & 7 \\ \hline
6~jets (loose)  & 9   & 1.8  & 10		& SS 2$\ell$ + 1$^+$ B-jets & 8 & 0.5 & 11 \\ \hline
6~jets (medium) & 7   & 1.7  & 10		& SS 2$\ell$ + 3$^+$ B-jets & 4 & 0.3 & 7 \\ \hline
6~jets (tight)  & 9   & 1.6  & 9 		&\multicolumn{4}{|c|}{} \\ \hline
\end{tabular}
\caption{\textbf{Results of Relevant ATLAS SUSY Searches}. The reported number of observed events, 95\% upper limit on the cross-section ($\sigma_{\rm BSM}$), and 95\% upper limit on the number of events ($N_{\rm BSM}$) for beyond the Standard Model contributions to the signal.}
\label{ATLAS}
\end{center}
\end{table}
%-------------------------------------------------------------------------------------

To date, none of the searches for supersymmetric particles has yielded a signal strength that is inconsistent with the background-only hypothesis. For all but one of the analyses we consider, the event rates were large enough to be able to establish an upper bound to the number of events, and effective cross-section, for contributions beyond that of the Standard Model at the 95\% level. The reported number of observed events, and upper limit on the cross-section ($\sigma_{\rm BSM}$) and number of events ($N_{\rm BSM}$), for these signal regions are given in Table~\ref{ATLAS}. In the absence of a proper treatment of backgrounds, we will utilize the reported $N_{\rm BSM}$ number to determine if a model point would have been detected in a given channel. The exceptional case was the same-sign dilepton analysis without b-tagged jets, where the observed number of events was very low: one event in the $e\,e$ and $\mu\,\mu$ channel, and two in the $e\,\mu$ channel. We will therefore consider a signal of three times this amount as an observable signal for the purposes of comparing a theoretical prediction to the data.

%-----------------------------------------------------------------------------------
\subsection{The BGW Model at ATLAS}

%%%%%%%%%%%%%%%%%% BENCHMARKS: GROSS LHC PHENO %%%%%%%%%%%%%%%%%%%%%%%%%
\begin{table}[t]
\begin{center}
\begin{tabular}{|c||c|c||c|c|c||c|c|c||c|c|c|c|c|} \hline
 &  \multicolumn{2}{|c||}{BGW Parameters} & \multicolumn{3}{|c||}{Key Masses (GeV)} & \multicolumn{3}{|c||}{Production} & \multicolumn{5}{|c|}{Gross Properties}\\ \hline
Point & \parbox{1.0cm}{$\beta_+$} & \parbox{1.0cm}{$m_{3/2}$} & \parbox{0.9cm}{$m_{\tilde{g}}$} & \parbox{0.9cm}{$m_{\tilde{N}_1}$} & \parbox{0.9cm}{$\Delta m$} & \parbox{1.5cm}{$\sigma_{\rm SUSY}^{\rm tot}$ (fb)}  & \parbox{1.1cm}{N($\tilde{g}\tilde{g}$)} & \parbox{1.1cm}{N($\tilde{\chi}\tilde{\chi}$)} & \parbox{1.3cm}{Trig. Eff.} & \parbox{1.1cm}{$\overline{N}_{\rm jet}$} & \parbox{1.1cm}{$\overline{N}_{\ell}$} & \parbox{1.1cm}{$N_{1b}$} & \parbox{1.1cm}{$N_{2b+}$} \\ \hline \hline
% 51985
A & 9 & 9425 & 498 & 246 & 18 & 2514 & 12917 & 1664 & 74.4\% & 7.5 & 0.7 & 12514 & 3858 \\ \hline
% 227445
B & 10 & 10465 & 628 & 266 & 17 & 633.2 & 2556 & 1117 & 71.1\% & 4.5 & 0.7 & 14321 & 10838 \\ \hline
% 217701
C & 11 & 10360 & 699 & 309 & 36 & 292.7 & 1155 & 543 & 71.8\% & 7.5 & 1.1 & 14185 & 11045 \\ \hline
% 218646
D & 12 & 10828 & 808 & 340 & 41 & 123.8 & 375 & 343 & 26.4\% & 7.2 & 1.1 & 11637 & 8963 \\ \hline
% 231808
E & 13 & 11175 & 913 & 369 & 44 & 63.0 & 135 & 230 & 58\% & 6.9 & 1.1 & 9720 & 6633 \\ \hline
% 225426
F & 14 & 11870 & 1050 & 366 & 17 & 48.6 & 38 & 244 & 39.4\% & 5.6 & 0.7 & 5206 & 2663 \\ \hline
% 218022
G & 15 & 11649 & 1114 & 422 & 46 & 23.8 & 22 & 116 & 49.0\% & 6.2 & 1.0 & 6518 & 3170\\ \hline
% 212095
H & 18 & 11994 & 1392 & 492 & 43 & 9.1 & 2 & 51 & 43.7\% & 5.8 & 0.8 & 5448 & 1269 \\ \hline
% 222261
I & 24 & 11983 & 1866 & 607 & 37 & 2.6 & 0 & 15 & 14.6\% & 5.3 & 0.7 & 4178 & 532 \\ \hline
% 232164
J & 33 & 11457 & 2437 & 707 & 16 & 1.1 & 0 & 6 & 9.3\% & 4.5 & 0.4 & 1924 & 175 \\ \hline
% 223734
K & 42 & 11369 & 3031 & 637 & 5 & 2.1 & 0 & 12 & 16.3\% & 3.9 & 0.2 & 774 & 24 \\ \hline
% 106272
L & 60 & 8279 & 3099 & 869  & 22 & 0.3 & 0 & 1 & 21.3\% & 4.5 & 0.5 & 1963 & 95 \\ \hline
% 233646
M & 60 & 10502 & 3877 & 577 & 3 & 3.8 & 0 & 22 & 15.4\% & 3.8 & 0.1 & 691 & 19 \\ \hline
% 4253
N & 81 & 9996 & 4800 & 498 & 3 & 8.4 & 0 & 49 & 15.1\% & 3.8 & 0.1 & 683 & 9 \\ 
\hline
\end{tabular}
\caption{\textbf{Global Properties of Benchmark Points for LHC Searches}. Superpartner production at the LHC for the BGW model is dominated by pair production of gluinos and electroweak gauginos. The mass of the gluino and lightest neutralino are given in~GeV, as well as the mass gap $\Delta m$ between the lightest neutralino and the lightest chargino. The overall production cross-section is given for $\sqrt{s} = 8\,{\rm TeV}$, and the number of gluino pairs and electroweak gaugino pairs produced is given for an integrated luminosity of 5.8~fb$^{-1}$. Triggering efficiency is estimated from a sample of 50,000 events, passed through the {\tt PGS} level-one trigger selection. Also listed is the mean number of jets ($\overline{N}_{\rm jet}$) and mean number of leptons ($\overline{N}_{\ell}$) across the 50,000 simulated events, as well as the number of these events with precisely one b-tagged jet ($N_{1b}$) or two or more b-tagged jets ($N_{2b+}$).}
\label{grossLHC}
\end{center}
\end{table}
%%%%%%%%%%%%%%%%%% BENCHMARK TABLE %%%%%%%%%%%%%%%%%%%%%%%%%

Looking at the benchmarks in Table~\ref{masstable}, it is clear that only about half of the points could reasonably be expected to be probed at the LHC with a center-of-mass energy of $\sqrt{s}=8\,{\rm TeV}$, even with the full 21.7~fb$^{-1}$ dataset. This is verified in Table~\ref{grossLHC}, where certain global properties of the benchmark points are given which are relevant for supersymmetry searches at the LHC. The most important of these is the gluino mass, which increases roughly linearly with the value of $\beta_+$, particularly for low values of this parameter. Gluino pair production is the dominant SUSY production channel up to  $m_{\tilde{g}} \simeq 800\,{\rm GeV}$, at which point production of electroweak gauginos (neutralinos and charginos) becomes dominant. 
This cross-over occurs at slightly higher gluino masses than in minimal supergravity as the lowest-lying eigensates of the chargino and neutralino mass matrices tend to be about twice as heavy as in an analogous mSUGRA model. In the table, the overall SUSY production cross section is given at $\sqrt{s} = 8\,{\rm TeV}$, while the number of gluino pair production events and electroweak gaugino pair production events is given at a fixed integrated luminosity of 5.8~fb$^{-1}$. We note that squarks and sleptons are never produced at the LHC at $\sqrt{s} = 8\,{\rm TeV}$ for any of these benchmark points.

For all of these points the value of $\mu$ is well below one~TeV, which imparts a substantial Higgsino component to the LSP. Consequently, the mass gap between the LSP and the lightest chargino (or nearly equivalently, between the second lightest neutralino and the LSP), indicated by the value of $\Delta m$ in Table~\ref{grossLHC}, tends to be small, but still sufficiently large to occasionally produce energetic jets and leptons. 
Nevertheless, when production of electroweak gauginos begins to dominate the overall triggering efficiency, as estimated by the level-one trigger selection in {\tt PGS4}, tends to drop significantly. The rather low mean lepton count -- generally at or less than one high-$p_T$ lepton per event -- and soft jet production makes these events both difficult to trigger upon and difficult to separate from Standard Model backgrounds, despite the reasonable number of events produced for  $\beta_+ \lappeq 15$ in 5.8~fb$^{-1}$ of data. We note that the level-one trigger in {\tt PGS} can be satisfied by a single inclusive jet with $p_T > 400\,{\rm GeV}$, or by a jet with $p_T > 180\,{\rm GeV}$ combined with $E_T^{\rm miss} > 80\,{\rm GeV}$. In general it is the large jet-$p_T$ requirement that adversely effects the trigger efficiency for the BGW model. In the analysis that follows we will \textbf{not} impose any trigger requirements, but assume all produced signal events are recorded.

The low lepton multiplicity immediately suggests that multi-lepton signatures are not effective discovery channels for the BGW~model, with the possible exception of same-sign dilepton events arising from gluino pair production events. The small mass gaps $\Delta m$ between kinematically accessible electroweak gaugino states implies that there should be no on-shell Z-bosons, thereby eliminating another possible channel from the analysis. The overall jet multiplicity (defined as jets with $p_T > 40\,{\rm GeV}$) is in the 5-8~jet range for most of the cases with $m_{\tilde{g}} \leq 2\,{\rm TeV}$, with the exception being point~B for which the branching fraction BR($\tilde{g} \to g\,\wtd{N}_i$) approaches 50\%. Generally at least one, and often more than one, of these jets is identified as arising from a bottom quark.

%%%%%%%%%%%%%%%%%% BENCHMARKS: SIGNAL REGION COUNTS %%%%%%%%%%%%%%%%%%%%%%%%%
\begin{table}[t]
\begin{center}
\begin{tabular}{|c||c|c||c|c|c|c|c|c|c|c||c|c||c|c||c|c|c|}
\multicolumn{3}{c}{ } & \multicolumn{8}{c}{Low Multiplicity Jets} & \multicolumn{7}{c}{Leptonic Channels} \\ \hline
\multicolumn{3}{|c||}{ } & \multicolumn{2}{|c|}{2 Jets} & \multicolumn{2}{|c|}{3 Jets} & \multicolumn{3}{|c|}{4 Jets} &  & \multicolumn{2}{|c||}{1 Lepton} & \multicolumn{2}{|c||}{SS Dilepton}  & \multicolumn{3}{|c|}{SS 2$\ell$, B-Jets}\\ \hline
Point & \parbox{0.9cm}{$\beta_+$} & \parbox{0.9cm}{$m_{\tilde{g}}$} & \parbox{0.8cm}{M} & \parbox{0.8cm}{T} & \parbox{0.8cm}{M} & \parbox{0.8cm}{T} & \parbox{0.8cm}{L} & \parbox{0.8cm}{M} & \parbox{0.8cm}{T} & \parbox{0.8cm}{5J} & \parbox{0.8cm}{1$e$} & \parbox{0.8cm}{1$\mu$} & \parbox{0.8cm}{$e\,\mu$} & \parbox{0.8cm}{$\mu\,\mu$} & \parbox{0.8cm}{0b} & \parbox{0.8cm}{1b} & \parbox{0.8cm}{3b}\\ \hline \hline
% 51985
A & 9 & 498 & 24 & \textbf{9} & 32 & 6 & \textbf{101} & \textbf{27} & \textbf{5} & \textbf{6} & 2 & \textbf{12} & \textbf{6} & 1 & \textbf{24} & \textbf{15} & 2 \\ \hline
% 227445
B & 10 & 628 & 6 & 1 & 10 & 1 & 39 & 11 & 1 & 2 & 2 & \textbf{11} & \textbf{6} & 1 & \textbf{33} & \textbf{68} & \textbf{21} \\ \hline
% 217701
C & 11 & 699 & 4 & 1 & 6 & 1 & 23 & 7 & -- & 1 & 1 & \textbf{7} & \textbf{8} & \textbf{4} & \textbf{22} & \textbf{44} & \textbf{13} \\ \hline
% 218646
D & 12 & 808 & 2 & -- & 3 & -- & 13 & 3 & -- & -- & 1 & 4 & \textbf{6} & \textbf{3} & \textbf{12} & \textbf{33} & \textbf{9} \\ \hline
% 231808
E & 13 & 913 & 2 & -- & 2 & -- & 7 & 2 & -- & -- & 1 & 2 & 3 & 2 & \textbf{7} & \textbf{18} & 4 \\ \hline
% 225426
F & 14 & 1050 & 2 & -- & 2 & -- & 4 & 2 & -- & -- & -- & 2 & 1 & 1 & 2 & 6 & 1\\ \hline
% 218022
G & 15 & 1114 & 1 & -- & 1 & -- & 2 & 1 & -- & -- & -- & 1 & 1 & 1 & 2 & 4 & 1 \\ \hline
% 212095
H & 18 & 1392 & -- & -- & -- & -- & -- & -- & -- & -- & -- & -- & -- & -- & -- & 1 & -- \\ \hline \hline
\multicolumn{3}{|r||}{Observed}  & 111 & 10 & 106 & 7 & 156 & 31 & 1 & 5 & 10 & 4 & 2 & 1 & 5 & 8 & 4\\
\multicolumn{3}{|r||}{$N_{\rm BSM}$} & 34 & 9 & 44 & 7 & 66 & 18 & 3 & 6 & 10 & 6 & 6 & 3  & 7 & 11 & 7 \\ 
\hline
\end{tabular}
\caption{\textbf{Event Counts for BGW Benchmark Points at $\sqrt{s} = 8\,{\rm TeV}$ for Selected ATLAS Searches.} Signal events are displayed for selected ATLAS search channels described in Section~\ref{searches}. Chosen channels were those in which one or more benchmark point generated a signal comparable in size to the 95\% confidence level upper bound on the number of signal events ($N_{\rm BSM}$), as reported by the ATLAS collaboration. These channels include most of the low-multiplicity multijet channels, the single lepton and same-sign dilepton channels, and (especially) the same-sign dilepton channel with accompanying b-tagged jets. The last three columns represent a simulated integrated luminosity of 20.7~fb$^{-1}$, while the others involve 5.8~fb$^{-1}$. Also given is the number of events observed by ATLAS in each channel, corresponding to the entries in Table~\ref{ATLAS}. Table entries in boldface indicate a channel which would have produced a discovery for that point.}
\label{benchSR}
\end{center}
\end{table}
%%%%%%%%%%%%%%%%%% BENCHMARK TABLE %%%%%%%%%%%%%%%%%%%%%%%%%

Table~\ref{benchSR} gives the number of events satisfying the ATLAS search criteria in our simulated data for benchmark points A-H. For points with heavier gluinos no events would have been observed in any of the ATLAS searches described in Section~\ref{searches}. We display only those channels for which one or more benchmark points generated a signal comparable in size to the 95\% confidence level upper bound on the number of signal events ($N_{\rm BSM}$) reported by the ATLAS collaboration. We remind the reader that the last three columns are scaled to an integrated luminosity of 20.7~fb$^{-1}$, while the others are scaled to an integrated luminosity of 5.8~fb$^{-1}$. Also given is the number of events observed by ATLAS in each channel, corresponding to the entries in Table~\ref{ATLAS}. 

For each benchmark, if a particular channel resulted in more signal events than the reported value of $N_{\rm BSM}$, the corresponding number of events is entered into Table~\ref{benchSR} in boldface. Thus we see that the overall reach across all channels is somewhere at, or just below, 1~TeV in the gluino mass. This is comparable to, but slightly weaker than, the reach in the minimal supergravity scenario. This statement is true both globally -- across all search channels -- and for each channel considered individually. The reach is best in the same-sign dilepton channels which often arise in gluino pair production events, and for which the Standard Model background rates are low enough to produce a rather small value for $N_{\rm BSM}$. In terms of the theoretical parameter space, this immediately implies that models with $\beta_+ \leq 12$, and a large fraction of the parameter space with $\beta_+ = 13$, are now no longer viable in light of ATLAS supersymmetry searches. Among these points are all the cases in which the LSP has a sizable wino component.

%=(8)=============== NGW vs mSUGRA: Meff ==================
\begin{figure}[th]
\begin{center}
\includegraphics[width=0.7\textwidth]{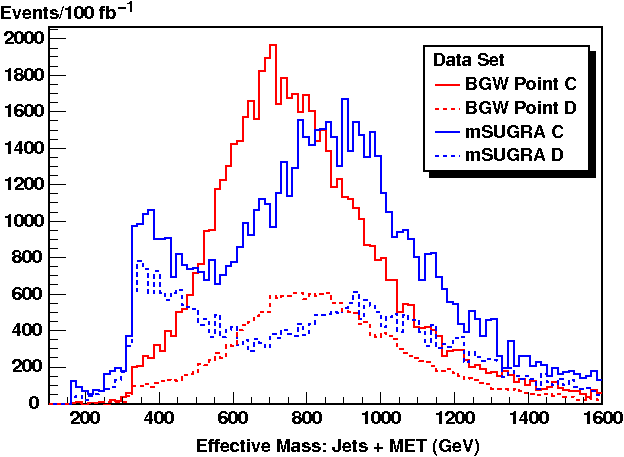}
\caption{\textbf{Effective Mass Distribution of Jets Plus $E_T^{\rm miss}$.} The distribution in the effective mass variable formed from $E_T^{\rm miss}$ plus the jet-$p_T$ for all jets with $p_T > 40\,{\rm GeV}$ is shown for BGW~benchmarks~C (with $m_{\tilde{g}} = 699\,{\rm GeV}$) and~D (with $m_{\tilde{g}} = 808\,{\rm GeV}$) and their mSUGRA analogs. The models with 700~GeV gluinos are the solid lines, while the dashed lines are the models with 800~GeV gluinos. %Note that the mSUGRA models (blue histograms) show evidence of the electroweak gaugino production in the low effective mass bins, which is mostly absent in the BGW models (red histograms).
}
\label{plot:meff}
\end{center}
\end{figure}
%==============================================================================

The combination of low jet-$p_T$, and jet multiplicities typically below 8~jets, resulted in negligible numbers of events in the high-multiplicity multijet channels, and the low-multiplicity jets plus $E_T^{\rm miss}$ channels fared little better. The reach in this set of channels was no better than roughly 500~GeV in the gluino mass. This is to be compared with a reported reach of up to 950~GeV in the mSUGRA/CMSSM paradigm, or 1100~GeV in the `simplified model' with very heavy scalars and a massless LSP. Of course, the previous numbers involved a combination across all channels in the low-multiplicity multijet search, but there is still a sizable difference between the reach for these models and the BGW~scenario. That the reach is higher when the lightest neutralino is assumed to be massless is easily understood; the resulting jets in the gluino decays are able to carry more $p_T$ from the increased phase space, and thereby generate a larger value of the effective mass variable for the same value of the gluino mass. To reduce the sizeable Standard Model background in these channels, the effective mass cut was set at rather substantial values: at least one~TeV in the `loose' variants of each channel, and generally much higher for the `medium' and `tight' variants of these channels. 

The BGW~model works in the opposite direction from the simplified models. The mass difference between the gluinos and the low-lying electroweak gauginos is smaller than in the mSUGRA model, and significantly so. While not quite to the extreme level as some `compressed spectrum' models considered in the literature~\cite{LeCompte:2011cn,Rolbiecki:2012gn}, the mass differences can have a real impact when the effective mass threshold is set very high. Consider benchmark points~C and~D from Tables~\ref{masstable}, \ref{grossLHC}~and~\ref{benchSR}, which have gluinos of roughly 700~and 800~GeV, respectively. A point in mSUGRA with a gluino of the equivalent mass should be discoverable at ATLAS already with 5.8~fb$^{-1}$ of data, yet neither of the two BGW~points generates a strong signal in the multijets channel. To understand why, we studied two look-alike points in minimal supergravity for BGW~points C~and~D, with $m_0$ set to the value of $m_{3/2}$ from Table~\ref{masstable}, $A_0=0$, $\tan\beta=42.5$ and $m_{1/2}$ chosen so as to generate a gluino of precisely the same mass as the BGW~analog. The number of gluino pairs produced at a given value of integrated luminosity is therefore precisely the same between the look-alikes.

The total production cross-section for superpartners was up to three times larger for the mSUGRA analogs given their much lighter electroweak gauginos. Therefore the fraction of events involving gluino pairs at a given integrated luminosity was smaller for the mSUGRA points: 25\% and 16\% for mSUGRA~C and mSUGRA~D versus 68\% and 52\% for BGW point~C and BGW point~D. However, since electroweak gaugino production produces quiet events in both models, they tend to have lower values of $E_T^{\rm miss}$, so after requiring $E_T^{\rm miss} > 160\,{\rm GeV}$ both points have roughly the same number of events. 

The distribution of the effective mass, formed from the jets with $p_T > 40\,{\rm GeV}$ and the missing transverse energy, is shown in Figure~\ref{plot:meff} for the BGW benchmarks~C and~D, as well as the two mSUGRA look-alikes. The cases with $m_{\tilde{g}} \simeq 700\,{\rm GeV}$ are given by the solid lines. The cases with $m_{\tilde{g}} \simeq 800\,{\rm GeV}$ are given by the dashed lines. Clearly, the mSUGRA models give generally larger typical values for the effective mass, with the cross-over occurring near $m_{\rm eff} \simeq m_{\tilde{g}}$. Demanding large values of this variable therefore favors the mSUGRA models considerably. Note that the mSUGRA models (blue histograms) show evidence of the electroweak gaugino production in the low effective mass bins, which is mostly absent in the BGW models (red histograms).

By the end of the summer in~2012 two crucial pieces of information were well established in ATLAS data: (1) a Higgs-like boson consistent with the Standard Model with mass near 125-126~GeV had been discovered, and (2) no sign of supersymmetry had been detected in the standard search channels, in roughly 6~fb$^{-1}$ of data. Far from being mutually inconsistent, these results are exactly what one might predict in supersymmetric theories with very large scalar masses and gluinos with masses around 1~TeV. In these scenarios in which squarks are quite heavy, gluino pair production becomes the only hope for discovery, and the branching fractions for gluino decays into third-generation quarks can be sizeable. The BGW~model is precisely such a theory.

The ATLAS collaboration has therefore begun designing searches with selection criteria optimized to this paradigm. The same-sign dilepton search with accompanying b-tagged jets is one such analysis which is quite effective at pushing the reach in the BGW~model back up to near the 1~TeV mark in gluino masses. The $0b$ signal region, with a veto on b-jets, is similar to the inclusive same-sign dilepton search with 5.8~fb$^{-1}$, and signal rates for the BGW~benchmarks do tend to scale up with integrated luminosity proportionately. Note that the effective mass cut in the b-jet analysis is not a significant deviation from earlier analyses, since the signal requirements in the previous analysis already implied at least 390~GeV of effective mass from the two leptons, four jets (at $p_T>50\,{\rm GeV}$ per jet) and $E_T^{\rm miss}$.

%%%%%%%%%%%%%%%%%%%%%%%%%%%%%%%%%%%%%%%%%%%%%%%%%%%%%%%%%%%%%%%%%%%%%%%
\section{Prospects for the BGW Model at Future Detection Experiments}
\label{futureLHC}

\subsection{LHC Searches at $\sqrt{s}=8\,{\rm TeV}$ and $\sqrt{s}=13\,{\rm TeV}$}

The total integrated luminosity recorded by ATLAS at $\sqrt{s}=8\,{\rm TeV}$ in 2012 amounted to 21.7~fb$^{-1}$. We imagine, therefore, that earlier general-purpose SUSY search analyses performed at much lower integrated luminosities may be updated to reflect the full data set collected prior to the recent shutdown. Without a full background analysis it is impossible to precisely calculate the reach in the BGW~model with this projected data set, but we can indicate the number of pairs of gluinos and electroweak gauginos that can be expected to have been recorded prior to the shutdown. This information is summarized in Table~\ref{future}.

%%%%%%%%%%%%%%%%%% FUTURE LHC %%%%%%%%%%%%%%%%%%%%%%%%%
\begin{table}[t]
\begin{center}
\begin{tabular}{|c||c|c||c||c|c|c||c|c|c|} \hline
 &  \multicolumn{2}{|c||}{BGW Parameters} &  & \multicolumn{3}{|c||}{21.7 fb$^{-1}$ at $\sqrt{s}=8\,{\rm TeV}$} & \multicolumn{3}{|c|}{100 fb$^{-1}$ at $\sqrt{s}=13\,{\rm TeV}$} \\  \hline
Point & \parbox{1.0cm}{$\beta_+$} & \parbox{1.0cm}{$m_{3/2}$} &\parbox{0.9cm}{$m_{\tilde{g}}$} & \parbox{1.5cm}{$\sigma_{\rm SUSY}^{\rm tot}$ (fb)}  & \parbox{1.1cm}{N($\tilde{g}\tilde{g}$)} & \parbox{1.1cm}{N($\tilde{\chi}\tilde{\chi}$)} & \parbox{1.5cm}{$\sigma_{\rm SUSY}^{\rm tot}$ (fb)}  & \parbox{1.1cm}{N($\tilde{g}\tilde{g}$)} & \parbox{1.1cm}{N($\tilde{\chi}\tilde{\chi}$)} \\ \hline \hline
% 218646
D & 12 & 10828 & 808 & 123.8 & 1403 & 1284 & 864 & 68221 & 18156 \\ \hline
% 231808
E & 13 & 11175 & 913 & 63.0 & 506 & 861 & 426 & 29695 & 12891 \\ \hline
% 225426
F & 14 & 11870 & 1050 & 48.6 & 141 & 914 & 242 & 11007 & 13173 \\ \hline
% 218022
G & 15 & 11649 & 1114 & 23.8 & 81 & 435 & 143 & 7023 & 7266 \\ \hline
% 212095
H & 18 & 11994 & 1392 & 9.1 & 7 & 189 & 48 & 1194 & 3632 \\ \hline
% 222261
I & 24 & 11983 & 1866 & 2.6 & -- & 56 & 14 & 81 & 1333 \\ \hline
% 232164
J & 33 & 11457 & 2437 & 1.1 & -- & 24 & 6 & 4 & 642 \\ \hline
% 223734
K & 42 & 11369 & 3031 & 2.1 & -- & 46 & 10 & -- &  995 \\ \hline
% 106272
L & 60 & 8279 & 3099 & 0.3 & -- & 5 & 2 & -- & 205 \\ \hline
% 233646
M & 60 & 10502 & 3877 & 3.8 & -- & 82 & 15 & -- & 1547 \\ \hline
% 4253
N & 81 & 9996 & 4800 & 8.4 & -- & 183 & 30 & -- & 3001 \\ 
\hline
\end{tabular}
\caption{\textbf{LHC Superpartner Production Rates for BGW Benchmark Points}. Production cross sections at $\sqrt{s}=8\,{\rm TeV}$ and $\sqrt{s}=13\,{\rm TeV}$ are given for BGW~benchmark points D-N. Also shown are the number of gluino pairs and electroweak gaugino pairs produced in the 2012 data (21.7~fb$^{-1}$ at $\sqrt{s}=8\,{\rm TeV}$) and in 100~fb$^{-1}$ at $\sqrt{s}=13\,{\rm TeV}$.}
\label{future}
\end{center}
\end{table}
%%%%%%%%%%%%%%%%%% BENCHMARK TABLE %%%%%%%%%%%%%%%%%%%%%%%%%

As has been emphasized elsewhere~\cite{Baer:2011aa}, the kinematic reach for gluino pairs has largely been saturated. We do not expect the reach to increase by more than about 100~GeV in the gluino mass when multijet and single-lepton analyses are updated to incorporate the full $\sqrt{s}=8\,{\rm TeV}$ data. That is, we might expect $\beta_+ \leq 14$ to be probed in data already recorded, with an updated multijet search covering the region with $m_{\tilde{g}} \lappeq 800\,{\rm GeV}$.

After the shutdown, when the center-of-mass energy increases to $\sqrt{s}=13\,{\rm TeV}$, we expect much of the theoretically relevant parameter space to be probed immediately. In the first 100~fb$^{-1}$ after the shutdown we expect the reach to increase to at least $\beta_+ \leq 18$, and likely to $\beta_+ \leq 24$. At this point, most of the theoretically motivated values of $\beta_+$ from realistic orbifold and Calabi-Yau compactifications of heterotic string theory will be within reach. This presumed reach of about 1800~GeV in the gluino mass with 100~fb$^{-1}$ of data at $\sqrt{s}=13\,{\rm TeV}$ is consistent with previous studies~\cite{Baer:2012vr}. The ultimate reach could be increased significantly if electroweak gaugino production could be efficiently triggered upon and measured, given the reasonably large number of events produced for all BGW~benchmarks.  %\textbf{More here and CITES}.

\subsection{Direct Detection of Neutralino Dark Matter}

%%%%%%%%%%%%%%%%%% BENCHMARK TABLE %%%%%%%%%%%%%%%%%%%%%%%%%
\begin{table}[t]
\begin{center}
\begin{tabular}{|c||c|c||c|c|c|c||c||c|c||c|c||c|} \hline
 &  \multicolumn{2}{|c||}{ } &  \multicolumn{4}{|c||}{ } &  & \multicolumn{4}{|c||}{Recoil Events in LXe, 5-25~keV} & Monochromatic \\ 
 &  \multicolumn{2}{|c||}{BGW Parameters} & \multicolumn{4}{|c||}{LSP Properties} & &  \multicolumn{2}{|c||}{Unscaled} & \multicolumn{2}{|c||}{Scaled} & Gamma Flux \\  \hline
Point & \parbox{1.0cm}{$\beta_+$} & \parbox{1.0cm}{$m_{3/2}$} &\parbox{1.0cm}{$m_{\rm LSP}$} & \parbox{1.0cm}{B\%}  & \parbox{1.0cm}{W\%} & \parbox{1.0cm}{H\%} 
& \parbox{1.0cm}{$\Omega_{\chi} h^2$} & XE100 & LUX300 & XE100 & LUX300 & cm$^{-2}$s$^{-1}$ \\ \hline
% 51985
A & 9 & 9425 & 246 & 67.9\% & 18.2\% & 13.9\% & 0.018 & 3.8 & 14.9 & 0.6 & 2.3 & $6.13 \times10^{-12}$ \\ 
% 227445
B & 10 & 10465 & 266 & 28.7\% & 20.9\% & 50.4\% & 0.010 & 11.4 & 44.7 & 1.0 & 3.9 & $1.71 \times10^{-12}$ \\ 
% 217701
C & 11 & 10360 & 309 & 84.8\% & 5.4\% & 9.8\% & 0.114 & 1.3 & 5.2 & 1.3 & 5.2 & $8.91 \times10^{-13}$ \\ 
% 218646
D & 12 & 10828 & 340 & 81.4\% & 5.1\% & 13.5\% & 0.115 & 1.6 & 6.3 & 1.6 & 6.3 & $3.89 \times10^{-13}$ \\ 
% 231808
E & 13 & 11175 & 369 & 78.5\% & 4.6\% & 16.7\% & 0.115 & 1.7 & 6.8 & 1.7 & 6.8 & $2.02 \times10^{-13}$ \\ 
% 225426
F & 14 & 11870 & 366 & 18.4\% & 8.4\% & 73.2\% & 0.020 & 5.0 & 19.7 & 0.9 & 3.4 & $2.67 \times10^{-13}$ \\ 
% 218022
G & 15 & 11649 & 422 & 73.1\% & 3.8\% & 23.1\% & 0.113 & 1.8 & 7.3 & 1.8 & 7.1 & $8.27 \times10^{-14}$  \\ 
% 212095
H & 18 & 11994 & 492 & 66.4\% & 2.9\% & 30.7\% & 0.114 & 1.8 & 6.9 & 1.7 & 6.9 & $3.81 \times10^{-14}$ \\ 
% 222261
I & 24 & 11983 & 607 & 56.7\% & 1.9\% & 41.5\% & 0.118 & 1.5 & 5.8 & 1.5 & 6.0 & $1.87 \times10^{-14}$ \\ 
% 232164
J & 33 & 11457 & 707 & 16.4\% & 1.7\% & 81.8\% & 0.068 & 1.0 & 4.1 & 0.6 & 2.4 & $2.16 \times10^{-14}$ \\ 
% 223734
K & 42 & 11369 & 637 & 1.1\% & 0.6\% & 98.3\% & 0.045 & 0.2 & 0.8 & 0.1 & 0.3 & $3.91 \times10^{-14}$ \\ 
% 106272
L & 60 & 8279 & 869 & 28.0\%  & 0.9\% & 71.2\% & 0.120 & 0.9 & 3.5 & 0.9 & 3.6 & $8.92 \times10^{-15}$ \\ 
% 233646
M & 60 & 10502 & 577 & 0.3\% & 0.2\% & 99.5\% & 0.036 & 0.1 & 0.3 & -- & 0.1 & $3.59 \times10^{-14}$ \\ 
% 4253
N & 81 & 9996 & 498 & 0.1\% & 0.1\% & 99.7\% & 0.027 & 0.0 & 0.2 & -- & -- & $9.07 \times10^{-14}$ \\ 
\hline
\end{tabular}
\caption{\textbf{Dark Matter Phenomenology of Benchmark Points}. The mass of the lightest neutralino is given in~GeV for the fourteen benchmark points of Table~\ref{masstable}. The wavefunction composition of the LSP is given in terms of the bino, wino and Higgsino percentage. Thermal relic abundance is calculated using {\tt MicrOmegas}, as are the event rates in liquid Xenon and the flux of photons from the galactic center for the combined $\gamma \gamma$ and $\gamma Z$ monochromatic signals. The ``XE100'' signal represents the number of events in the reported Xenon100 exposure of 7636~kg-days, while ``LUX300'' represents the number of events in 300~days of exposure for the LUX 100~kg detector. For these event rates we show the expected events assuming a normalization of 0.3~GeV/cm$^3$ (unscaled), and a halo density scaled by the ratio of the predicted value of $\Omega_{\chi}h^2$ to the value extracted from WMAP~data (scaled).}
\label{DM}
\end{center}
\end{table}
%%%%%%%%%%%%%%%%%% BENCHMARK TABLE %%%%%%%%%%%%%%%%%%%%%%%%%

While it is reasonable to expect that all points in the parameter space of the BGW~model with $\beta_+ \leq 24$ will be probed at the LHC at $\sqrt{s} = 13\,{\rm TeV}$, higher values of the gluino mass may prove difficult to observe. 
%Indeed, one estimate puts the reach for $\sqrt{s}=14\,{\rm TeV}$ at only $m_{\tilde{g}} \lappeq 1800\,{\rm GeV}$ with 300~fb$^{-1}$ of data~\cite{Baer:2012vr}. 
But the prospects are somewhat brighter for these cases in terms of direct detection of neutralino dark matter. Table~\ref{DM} revisits the 14~benchmark points, but this time we focus on the properties of the lightest neutralino, and the implications for dark matter detection experiments. 

%=(9)=============== Fine scan, all constraints, sigmaSI vs mLSP ==================
\begin{figure}[t]
\begin{center}
\includegraphics[width=0.9\textwidth]{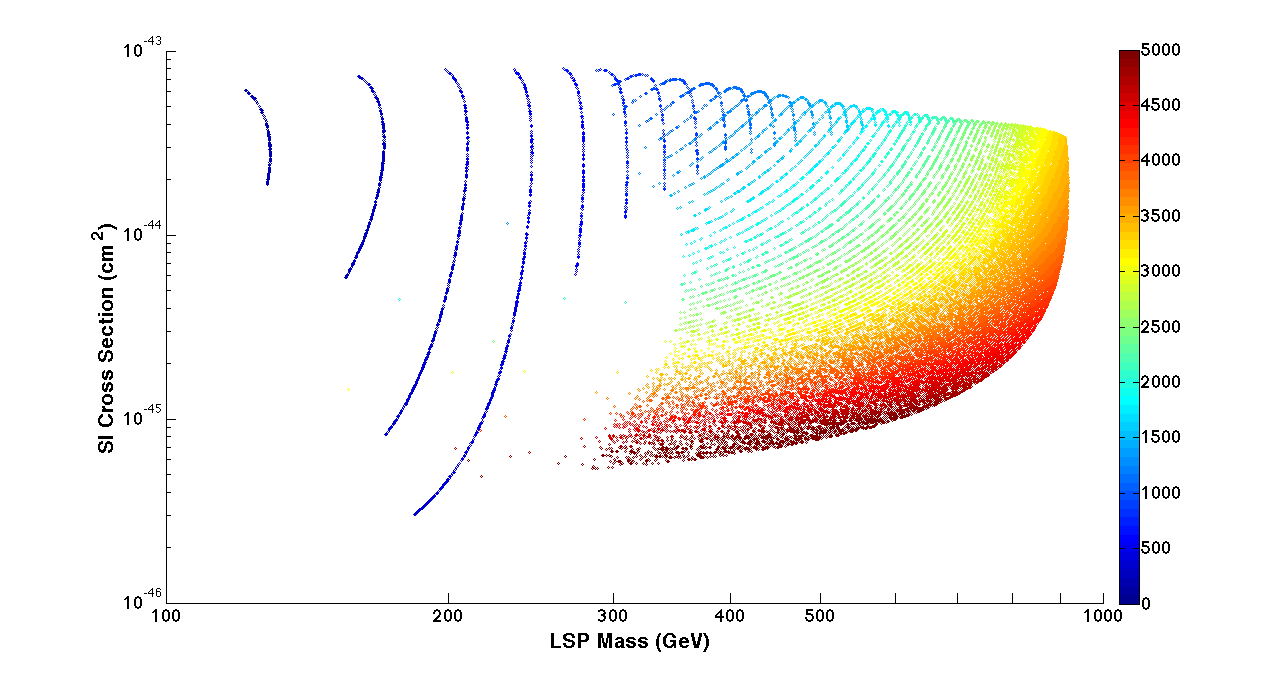}
\caption{\textbf{Spin-Indepenent Cross-section on Protons for Points in Figure~\ref{plot:fineGluino}.} Spin-independent cross-section on the proton for the~LSP versus LSP~mass for the models passing all constraints in Figure~\ref{plot:fineGluino}. The gluino mass (in units of GeV) is indicated by the color, as in Figure~\ref{plot:fineGluino}.}
\label{plot:fineSigma}
\end{center}
\end{figure}
%==============================================================================

Restricting our attention to cases which would not have been detected at the LHC implies a relatively heavy LSP neutralino: generally heavier than 300-350~GeV in mass. Many of the benchmark points generate a thermal relic abundance for the LSP commensurate with (or even slightly larger) than the central value extracted from WMAP~measurements of the cosmic microwave background. These points generally have a mixed LSP, split between bino and Higgsino components. Such neutralinos are well known to have a relatively large interaction cross-section for spin-independent elastic collisions with nucleons. Heavier mass cases tend to be almost exclusively Higgsino-like, with relatively smaller interaction cross-sections.

The points which passed all phenomenological constraints for $\tan\beta=42.5$, represented graphically by the points in Figure~\ref{plot:fineGluino}, were passed to {\tt MicrOmegas} where the spin-independent cross-section for elastic scattering from protons was computed. The distribution of these points versus the LSP~mass is shown in Figure~\ref{plot:fineSigma}. As before, the gluino mass is indicated in GeV by the color code shown on the right of the plot.
The striation of the plot arises both from the logarithmic scale of the horizontal axis and from the discrete nature of the input variable $\beta_+$. As this value increases, for a fixed value of $m_{3/2}$, the overall mass scale of the gauginos increases as well. The blue lines at low values of the LSP~mass correspond to small values of $\beta_+$. For these cases there is a wide variation in the composition of the LSP, from bino-like to wino-like, as the gravitino mass is varied. This results in a large range of possible spin-independent cross-sections. Note that these blue lines at low values of $\beta_+$ would already have been ruled out by direct searches for gluinos at the~LHC. At higher values of $\beta_+$ the distribution begins to be more continuous, and the LSP is more Higgsino-like across the range of allowed gravitino masses. 
The highest cross-sections correspond to model points with $10 \lappeq \beta_+ \lappeq 36$, where the LSP is a mixed neutralino dominated by the Higgsino component. Such states are known to have the largest cross-section for direct detection at heavy target experiments such as those based on liquid xenon~\cite{Baer:2006te}. In this case, these are also precisely the points with gluinos in a mass range where pair production at the LHC can be substantial.

Last summer, the Xenon100 experiment released the results of 225~days of exposure of their 34~kg liquid xenon detector~\cite{Aprile:2012nq}. Recoils were counted with energies between 6.6~and 30.5~keV (electron-equivalent). The detector observed two candidate events within the signal region, which was consistent with a background expectation of $1.0\pm 0.2$ events in the same 7636.4~kg-days of exposure. This translates into an upper bound on the spin-independent elastic cross-section for neutralino scattering on protons of $\sigma_{p}^{\rm SI} \leq 2.0 \times 10^{-45}\,{\rm cm^2}$, for a neutralino of mass 55~GeV. A large fraction of the surviving points of the BGW~parameter space have elastic scattering cross-sections well in excess of this limit -- including all of the points with $\beta_+ \leq 36$ with gluino masses above 800~GeV. However, the cross-section limits are significantly weaker for larger mass neutralinos: all of the points with $m_{\tilde{g}} \geq 800\,{\rm GeV}$ have cross-sections no larger than one-sigma above the central value of the upper bound for $m_{\rm LSP} \geq 300\,{\rm GeV}$. 

It is therefore more appropriate to compute the number of recoil events expected, per kilogram-day of exposure, in liquid xenon across a certain recoil energy window. We have chosen to compute the rate in the energy range 5-25~keV (electron equivalent). In Table~\ref{DM} we give the calculated number of events in 7636.4~kg-days at the Xenon100 experiment for the fourteen benchmark points. Points~A, B~and~F would have produced a signal larger than that actually observed by the experiment, though points~A and~B are already excluded by direct searches at the~LHC. Yet even this number is deceptive. To compute it we assumed a local halo density of $\rho = 0.3 \,{\rm GeV/cm^3}$ for these neutralinos, which best fits rotation models for objects in the Milky Way halo. In the case of point~F, the relatively high wino content results in a relic abundance roughly six times smaller than the measured abundance from~WMAP. This is not necessarily inconsistent, though it would tend to require some sort of non-thermal production of neutralinos in the early universe to reconcile the two values. Alternatively, if dark matter consists of multiple components, one might instead scale the event rate by this factor of six to account for the reduced flux of neutralino WIMPS in the detector. This `scaled' event count is also given in Table~\ref{DM}. Having performed this scaling, none of our benchmark points would have been inconsistent with the results from Xenon100.

\begin{figure}[t]
\begin{center}
\includegraphics[width=0.9\textwidth]{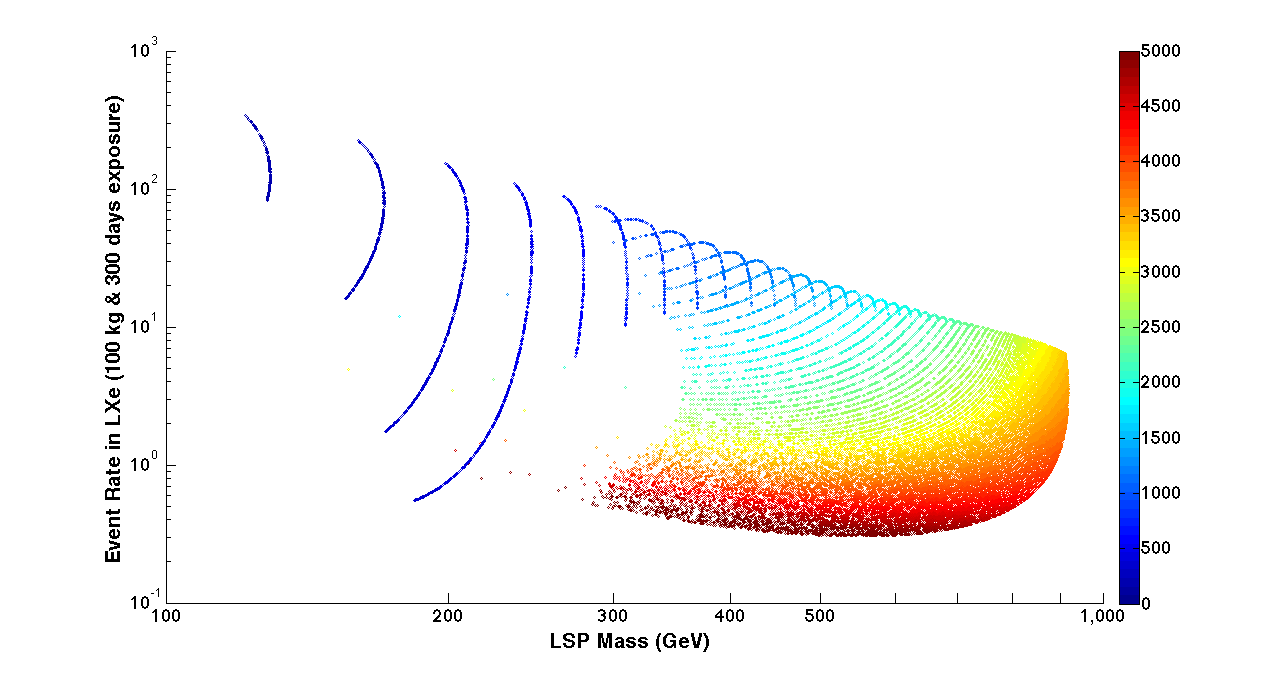}
\caption{\textbf{Event Rate in 300~Days Exposure at LUX.} Number of recoil events between 5~and 25~keV (electron equivalent) of recoil energy in liquid xenon, normalized to 300~days of 100~kg fiducial target. The values assume an abundance of neutralinos in the Milky Way halo equivalent to 0.3~GeV/cm$^3$. Points correspond to those of Figure~\ref{plot:fineSigma}, with the gluino mass (in units of GeV) indicated by the color, as in Figure~\ref{plot:fineGluino}.}
\label{plot:fineRate1}
\end{center}
\end{figure}

\begin{figure}[t]
\begin{center}
\includegraphics[width=0.9\textwidth]{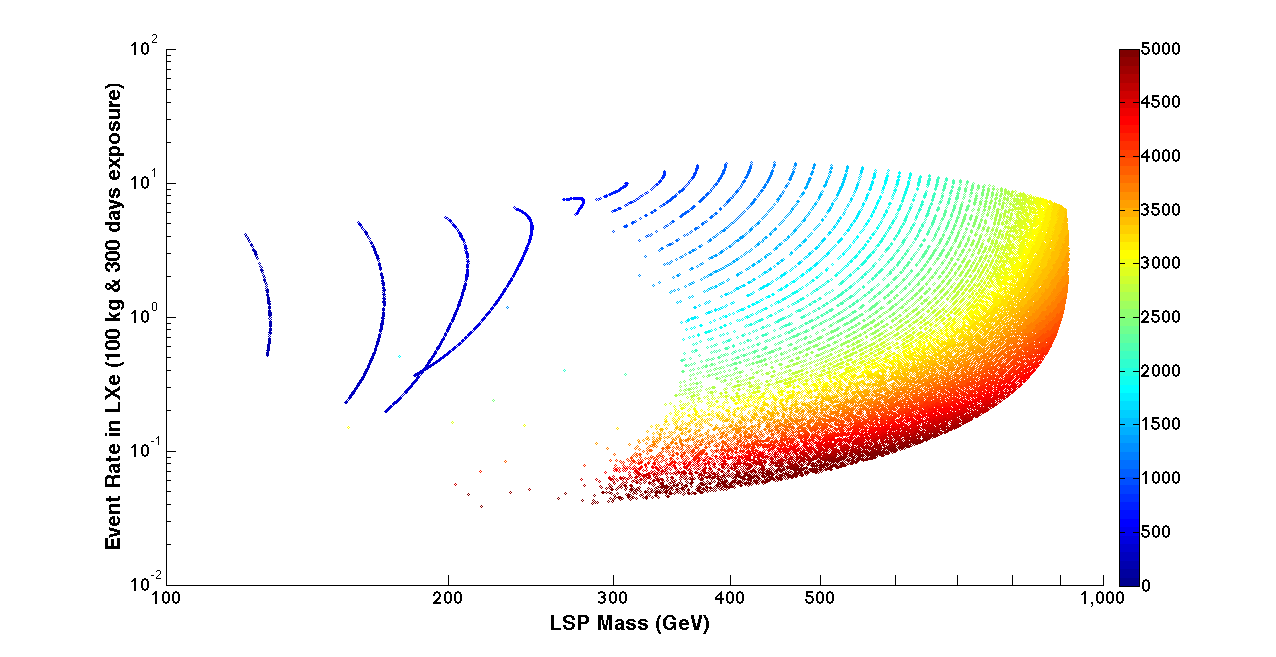}
\caption{\textbf{Event Rate in 300~Days Exposure at LUX, Scaled.} Number of recoil events between 5~and 25~keV (electron equivalent) of recoil energy in liquid xenon, normalized to 300~days of 100~kg fiducial target. The values have been scaled by the ratio of $\Omega_{\chi} h^2$ predicted by thermal production to the WMAP~abundance. Points correspond to those of Figure~\ref{plot:fineSigma}, with the gluino mass (in units of GeV) indicated by the color, as in Figure~\ref{plot:fineGluino}.}
\label{plot:fineRate2}
\end{center}
\end{figure}

The LUX experiment~\cite{Akerib:2012ys} has now been installed in its deep underground site and should release its first underground data this year. The LUX~experiment involves a 100~kg fiducial target mass, and the number of expected events in 300~days of exposure is given in Table~\ref{DM} both with, and without, scaling the event rate by the relative dark matter relic abundance. The LUX~collaboration intends to reach a background level of less than one event per 300~days of exposure~\cite{Akerib:2012ys,Akerib:2012ak}, suggesting that all but three of the benchmark points should give a signal in this level of data collection. This is even after re-scaling the data to account for lower thermal relic abundances. The event rate in 30,000~kg-days of exposure in liquid xenon is given for the entire set of BGW~points with $\tan\beta=42.5$ in Figures~\ref{plot:fineRate1} and~\ref{plot:fineRate2}. Figure~\ref{plot:fineRate1} assumes an abundance of neutralinos in the Milky Way halo equivalent to 0.3~GeV/cm$^3$, while Figure~\ref{plot:fineRate2} scales the event rate by the ratio of $\Omega_{\chi} h^2$ predicted by thermal production to the WMAP~abundance. Note that the effect of this scaling is most pronounced for the very lightest gluino masses (which are mostly eliminated from the LHC~searches of the previous section) and the heaviest gluino masses. 

%=(10)=============== Fine scan, all constraints, sigmaSI vs mLSP ==================
%==============================================================================

%=(11)=============== Fine scan, all constraints, sigmaSI vs mLSP ==================
%==============================================================================

This correlation between dark matter direct detection and the gluino mass is not uncommon in string-motivated supersymmetric models~\cite{Feldman:2010uv}. In the case of the BGW~model, we can take $m_{\tilde{g}} \geq 800\,{\rm GeV}$ as a rough bound on the gluino mass arising from LHC~searches. From above, the gluino mass is everywhere less than 5~TeV in the allowed parameter space, but it is much lighter for the values of $\beta_+$ that correspond to realistic hidden sectors from the point of view of compactifications of heterotic string theory. If the hidden sector is no larger than $E_6$ than we must have $\beta_+ \leq 36$, in which case {\em the model predicts a cross-section which will be probed by the LUX experiment in 2013}. This statement assumes that all of the dark matter density inferred from the WMAP~experiment is made up of relic neutralinos. The statement continues to hold, however, for $\beta_+ \leq 24$ even after scaling the event rate to account for a lower thermal relic abundance. In either case, ton-scale liquid xenon detectors will certainly give a robust signal across all of these points. The event rate in Figures~\ref{plot:fineRate1} and~\ref{plot:fineRate2} are easily scaled to units of ton-years by multiplying the leftmost axis by a factor of ten. Xenon1T -- the one-ton extension to Xenon100 -- is due to begin taking data in 2015, at about the time that the LHC resumes proton collisions. For the case in which we do not scale event rates by the thermal relic abundance we find that {\em all parameter points of the BGW~model will give a signal in one ton-year of exposure}.

Finally, we should expect a corroborating signal from neutralino annihilation at the galactic center, in the form of monochromatic photons with energies very near the mass of the~LSP. Excluding points which would have given rise to signals at the LHC in analyses published thus far, we find that the lightest neutralino will always be well above 200~GeV in mass. This suggests that the Fermi satellite will not be sensitive to the high energy gamma rays arising from loop-induced annihilation processes. Instead we must turn to ground based atmospheric Cherenkov telescopes (ACTs), whose resolution will generally not be sufficient to distinguish between the $\gamma \gamma$ and $\gamma\,Z$ annihilation channels. We therefore give the combined flux of gamma-ray photons from the direction of the galactic center, using the NFW halo profile, in the final column of Table~\ref{DM}. Despite the loop-induced nature of these signals, the large Higgsino component to the~LSP can boost the signal to a range that might be detectable in the future. For example, benchmark points~D-N would give rise to a monochromatic gamma ray signal above the threshold of 310~GeV for detection at the HESS~experiment~\cite{Abramowski:2013ax}. Observations of the galactic center of 112~hours, released in~2012, put a 95\% confidence level upper limit on this flux of $1.14\times 10^{-11} {\rm cm^{-2}\,s^{-1}}$ for a 500~GeV signal, assuming a cone size of 1$^o$ opening angle about the galactic center. The HESS~II upgrade now taking data has an effective collection area four times the size of its predecessor, but even after 200~hours of observation of the galactic center the reach in the monochromatic photon flux only rules out models that are already eliminated by LHC~data. The proposed Cherenkov Telescope Array (CTA)~\cite{Consortium:2010bc}, however, should be able to observe monochromatic photon fluxes to the $10^{-14}\,{\rm cm^{-2}\, s^{-1}}$ level~\cite{Doro:2012xx}, thereby providing important cross-checks on the most theoretically motivated parts of the BGW~parameter space.

%%%%%%%%%%%%%%%%%%%%%%%%%%%%%%%%%%%%%%%%%%%%%%%%%%%%%%%%%%%%%%%%%%%%%%
\section{Conclusions}

Weakly-coupled heterotic string theory, with dilaton stabilization through non-perturbative corrections to the dilatonic action, was the first manifestation of a pattern of supersymmetry breaking that later came to be known as `mirage mediation'. Since the~BGW model first appeared nearly twenty years ago, the pattern has emerged in many other contexts, most notably Type~IIB string theory compactified on orientifolds in the presence of fluxes. Such constructions are some of the best-motivated, and most-studied, models of low-energy particle physics from string theory. 

In this paper we have begun the process of confronting string models with data by focusing on one of the simplest models to analyze. The theory can be described by the confinement of a single gauge group, whose beta-function coefficient is given by the parameter $\beta_+$, and an overall mass-scale given by the gravitino mass $m_{3/2}$. All of the other intricacies of the model are related to achieving vanishing vacuum energy at the minimum of the dilaton potential, cancelling divergences, and achieving weak coupling at the string scale. From the low-energy point of view, the model is effectively two-dimensional, plus the specification of $\tan\beta$ for electroweak symmetry breaking. In this sense the model is a concrete manifestation of the more general dilaton domination scenario.

Because the model is relatively simple it is highly constrained by the LHC~data. This is an unequivocally good thing, and should put to rest the notion that string phenomenology is not a legitimate way to test string theory. In fact the data collected by the LHC, as we enter the first shutdown period, has left a region of parameter space for the BGW~construction that makes some specific predictions. Assuming an MSSM~field content, and focusing on the parameter space with $\beta_+ \leq 36$, we can list these predictions as follows: (1) scalars are inaccessible to the LHC and any scalar-mediated processes (such as rare B-meson decays) will be consistent with the Standard Model, (2) gluinos are no heavier than 2900~GeV for $\beta_+ \leq 36$, and are less than 2100~GeV (and therefore accessible at $\sqrt{s} = 13\,{\rm TeV}$) for hidden sectors with $\beta_+ \leq 24$, as is very typical in the orbifold limit of compactifications, (3) the Higgs mass will be less than 127~GeV and the Higgs boson will be Standard Model like in its couplings, (4) The value of $\tan\beta$ will be large (typically $\tan\beta \geq 40$), and the inferred value of the $\mu$ parameter will be small, (5) there will be a collection of low-lying neutralino and chargino states with mass gaps between them at the 1-10\% level, (6) the neutralino LSP will have a large Higgsino component (and may be entirely all Higgsino), (7) evidence of neutralino dark matter should be discovered at the~LUX experiment within the first year, or two years, of data-taking -- perhaps even before the LHC~resumes taking data.

Much of the above is true for any model of supersymmetry breaking with the MSSM field content, simply by virtue of the rather large Higgs mass and the absence of a supersymmetric signal at the~LHC. But what distinguishes the~BGW model (and mirage models, generally) from most of the models being considered in the literature, is the relatively large mass of the neutralinos relative to the gluino. One can expect a lightest neutralino that will be approximately twice as massive as that in a minimal supergravity model that otherwise satisfies all the experimental constraints. This is important, since it allows us to make one more prediction that will help to distinguish the BGW~paradigm from the other models being considered: {\em the BGW~model will not give rise to a monochromatic gamma ray signal that can be observed by the Fermi-LAT satellite}. Thus, the reported line signal at about 130~GeV must be interpreted as poorly understood astrophysical backgrounds~\cite{Profumo:2012tr,Finkbeiner:2012ez}. Instead, the heavy neutralinos in the BGW~parameter space will give rise to such a signal which can only be observed by atmospheric Cherenkov telescopes with large effective areas and/or large exposure times on the galactic center.

If one or more of these predictions fail to come to pass, is this effective theory falsified? In brief, yes it is. But we hasten to add that this is the simplest manifestation of hidden sector gaugino condensation in heterotic string theory with K\"ahler stabilization of the dilaton. Variations on the model, still within the overarching context of the supergravity framework of Section~\ref{bgw}, are possible. For example, heterotic string theory always contains at least one $U(1)$ factor which is anomalous in the four-dimensional effective theory. If observable sector states carry charges under this anomalous $U(1)$ then additional D-term contributions, arising from the Green-Schwarz mechanism, can alter the pattern of scalar masses. Furthermore, we have assumed that the GS counterterm in~(\ref{LGS}), which cancels the sigma-model anomalies in the effective supergravity theory, is independent of the matter fields in the theory. This is usually not the case, as matter fields arising from various twisted sectors may have more complicated involvements in the GS~counterterm. Such additional contributions to scalar masses could reduce the mass of certain squarks and sleptons, and may even make them relevant for LHC~phenomenology.

To the extent that the BGW~model remains an example of the `mirage pattern' of gaugino masses, however, the bulk of the phenomenology described in this paper will continue to hold. However, models in which two gauge groups, with closely tuned beta-function coefficients, compete to drive supersymmetry breaking -- so-called racetrack models -- alter a very different set of possibilities. While multiple condensing gauge groups in the hidden sector is a generic outcome, the coincidence of very similar beta-function coefficients needed for realistic minima is not. In the~BGW context, the group with the largest beta-function coefficient dominates, and this drives the subsequent phenomenology. In non-generic outcomes we may expect some of the above predictions to be evaded. Nevertheless, the dramatic results from the~LHC are already putting extreme pressure on supersymmetric theories, most especially those with  high-scale motivation from string theories.

\section{Acknowledgements}
BK would like to thank Gregory Peim, Sujeet Akula and Baris Altunkaynak for technical assistance in the early stages of this work. BK and BDN are supported by the NSF under grant PHY-0757959. MKG is supported in part by the Director, Office of Science, Office of High Energy and Nuclear Physics of the U.S. Department of Energy under Contract DE-AC02-05CH11231, and in part by the National Science Foundation under grant PHY-0457315.

\section{Appendix}

In the approximation that generational mixing can be neglected, so
that only third-generation Yukawa couplings are relevant, we have
% Gammas
\begin{eqnarray}
(16\pi^2)\gamma_{Q_3}&=&\frac{8}{3}g_{3}^{2} +\frac{3}{2}g_{2}^{2}
+\frac{1}{30}g_{1}^{2} -\lambda_{t}^{2} -\lambda_{b}^{2} \nonumber
\\ (16\pi^2)\gamma_{U_3}&=&\frac{8}{3}g_{3}^{2}
+\frac{8}{15}g_{1}^{2} -2\lambda_{t}^{2} \nonumber
\\ (16\pi^2)\gamma_{D_3}&=&\frac{8}{3}g_{3}^{2}
+\frac{2}{15}g_{1}^{2} -2\lambda_{b}^{2} \nonumber
\\ (16\pi^2)\gamma_{L_3}&=&\frac{3}{2}g_{2}^{2}
+\frac{3}{10}g_{1}^{2} -\lambda_{\tau}^{2} \nonumber
\\ (16\pi^2)\gamma_{E_3}&=& \frac{6}{5}g_{1}^{2}
-2\lambda_{\tau}^{2}\nonumber
\\ (16\pi^2)\gamma_{H_u}&=&\frac{3}{2}g_{2}^{2}
+\frac{3}{10}g_{1}^{2} -3\lambda_{t}^{2} \nonumber
\\ (16\pi^2)\gamma_{H_d}&=&\frac{3}{2}g_{2}^{2}
+\frac{3}{10}g_{1}^{2} -3\lambda_{b}^{2} -\lambda_{\tau}^{2} ,
\label{actualgammas}
\end{eqnarray}
while the closely related $\wtd{\gamma}_i$ factors are given by
\begin{eqnarray}
(16\pi^2)\wtd{\gamma}_{Q_3}&=&\frac{8}{3}g_{3}^{4} +\frac{3}{2}g_{2}^{4}
+\frac{1}{30}g_{1}^{4} -\frac{g_s^2}{2}\left(\lambda_{t}^{2} -\lambda_{b}^{2}\right) \nonumber
\\ (16\pi^2)\wtd{\gamma}a_{U_3}&=&\frac{8}{3}g_{3}^{4}
+\frac{8}{15}g_{1}^{4} -\frac{g_s^2}{2}\left(2\lambda_{t}^{2}\right) \nonumber
\\ (16\pi^2)\wtd{\gamma}_{D_3}&=&\frac{8}{3}g_{3}^{4}
+\frac{2}{15}g_{1}^{4} -\frac{g_s^2}{2}\left(2\lambda_{b}^{2}\right) \nonumber
\\ (16\pi^2)\wtd{\gamma}_{L_3}&=&\frac{3}{2}g_{2}^{4}
+\frac{3}{10}g_{1}^{4} -\frac{g_s^2}{2}\left(\lambda_{\tau}^{2}\right) \nonumber
\\ (16\pi^2)\wtd{\gamma}_{E_3}&=& \frac{6}{5}g_{1}^{4}
-\frac{g_s^2}{2}\left(2\lambda_{\tau}^{2}\right) \nonumber
\\ (16\pi^2)\wtd{\gamma}_{H_u}&=&\frac{3}{2}g_{2}^{4}
+\frac{3}{10}g_{1}^{4} -\frac{g_s^2}{2}\left(3\lambda_{t}^{2}\right) \nonumber
\\ (16\pi^2)\wtd{\gamma}_{H_d}&=&\frac{3}{2}g_{2}^{4}
+\frac{3}{10}g_{1}^{4} -\frac{g_s^2}{2}\left(3\lambda_{b}^{2} -\lambda_{\tau}^{2}\right) ,
\label{gammatilde}
\end{eqnarray}

\end{document}

%% file: basic.tex
\DeclareMathAlphabet   {\mathsc}{OT1}{cmr}{m}{sc}

\def\[{\left [}
\def\]{\right ]}
\def\({\left (}
\def\){\right )}
\newcommand{\lang}{\left\langle}
\newcommand{\rang}{\right\rangle}
\newcommand{\lbr}{\left\{}
\newcommand{\rbr}{\right\}}
\newcommand{\oline}[1]{\overline{#1}}

\newcommand{\wtd}[1]{\widetilde{#1}}
\newcommand{\Lag}{\mathcal{L}}

\newcommand{\GS}       {\mathsc{gs}}
\newcommand{\PL}       {\mathsc{pl}}

\newcommand{\STR}      {\mathsc{str}}

\newcommand{\superint}{\int\diff^{4}\theta}

\newcommand{\Wa}{W_{\alpha}}

\newcommand{\Wc}{W^{\alpha}}

\newcommand{\lowest}{|_{\theta =\bar{\theta}=0}}

\newcommand{\baal}{b_{a}^{\alpha}}
\newcommand{\baaleff}{\(\baal\)_{\rm eff}}

\newcommand{\caleff}{\({c_{\alpha}}\)_{\rm eff}}
\newcommand{\hc}       {\mathrm{\; h.c. \;}}

\newcommand{\diff}{\mbox{d}}
\newcommand{\order}{\mathcal{O}}

\newcommand{\gappeq}{\mathrel{\rlap {\raise.5ex\hbox{$>$}}
{\lower.5ex\hbox{$\sim$}}}}
\newcommand{\lappeq}{\mathrel{\rlap{\raise.5ex\hbox{$<$}}
{\lower.5ex\hbox{$\sim$}}}}